\documentclass[lettersize,journal]{IEEEtran}

\usepackage{amsmath,amsfonts,graphics,amssymb,epsfig,subfigure,color,cite}                             
\usepackage{amsthm,textcomp}
\theoremstyle{plain}
\newtheoremstyle{mystyle}
  {0mm}
  {0mm}
  {}
  {4mm}
  {\bfseries}
  {:}
  { }
  {\thmname{#1}\thmnumber{ #2}\thmnote{ (#3)}}
\theoremstyle{mystyle}
\usepackage{bm}
\usepackage{array}
\usepackage{multirow}
\usepackage{enumerate}
\usepackage{algorithm}
\usepackage{algpseudocode}
\usepackage{adjustbox}
\usepackage{algcompatible}
\algblockdefx{FORP}{ENDFORP}[1]%
  {\textbf{for }#1 \textbf{do in parallel}}%
  {\textbf{end}}
\algblockdefx{FOR}{ENDFOR}[1]%
  {\textbf{for }#1 \textbf{do}}%
  {\textbf{end}}
\algblockdefx{noEnd}{noEndEnd}[1]
    {#1}
    {{\Statex}}
\algblockdefx{BULLET}{EndBULLET}{$\bullet$ }{}
\algnotext{EndBULLET}
\algblockdefx{IF}{ENDIF}[1]%
  {\textbf{if }#1 \textbf{then}}%
  {\textbf{end}}
\algnewcommand\algorithmicprocedure{\textbf{function}}
\algnewcommand\FUNC{\item[\algorithmicprocedure]}%
\algnewcommand\algorithmicendprocedure{\textbf{end function}}
\algnewcommand\ENDFUNC{\item[\algorithmicendprocedure]}%
 \usepackage{setspace}
\let\Algorithm\algorithm
\renewcommand\algorithm[1][]{\Algorithm[#1]\setstretch{1.4}}
\usepackage{float}
\usepackage{makecell}
\usepackage{graphicx}

\usepackage[caption=false,font=normalsize,labelfont=sf,textfont=sf]{subfig}
\usepackage[right=0.625in, left=0.625in, top=0.7in, bottom=1.0in]{geometry}

\newtheorem{thm}{Theorem}
\newtheorem{lem}{Lemma}
\newtheorem{cor}{Corollary}

\usepackage{bbm}

\floatname{algorithm}{Algorithm}

\newcommand{\argmax}{\operatornamewithlimits{argmax}}

\makeatletter
\newcommand{\vast}{\bBigg@{4.5}}
\newcommand{\Vast}{\bBigg@{7.5}}
\makeatother

\allowdisplaybreaks

\begin{document}
    \title{Prior-Aware Robust Beam Alignment \\ for  Low-SNR Millimeter-Wave Communications}
            \author{Jihun Park, Yongjeong Oh, Jaewon Yun, Seonjung Kim, and Yo-Seb Jeon 
    \thanks{Jihun Park, Yongjeong Oh, Jaewon Yun, Seonjung Kim, and Yo-Seb Jeon are with the Department of Electrical Engineering, POSTECH, Pohang, Gyeongbuk 37673, South Korea (e-mail: jihun.park@postech.ac.kr; yongjeongoh@postech.ac.kr; jaewon.yun@postech.ac.kr; seonjung.kim@postech.ac.kr; yoseb.jeon@postech.ac.kr).}
    }
    \vspace{-2mm}	
    
    \maketitle

    \begin{abstract} 

        This paper presents a robust beam alignment technique for millimeter-wave communications in low signal-to-noise ratio (SNR) environments. The core strategy of our technique is to repeatedly transmit the most probable beam candidates to reduce beam misalignment probability induced by noise. Specifically, for a given beam training overhead, both the selection of candidates and the number of repetitions for each beam candidate are optimized based on channel prior information. To achieve this, a deep neural network is employed to learn the prior probability of the optimal beam at each location. The beam misalignment probability is then analyzed based on the channel prior, forming the basis for an optimization problem aimed at minimizing the analyzed beam misalignment probability. A closed-form solution is derived for a special case with two beam candidates, and an efficient algorithm is developed for general cases with multiple beam candidates. Simulation results using the DeepMIMO dataset demonstrate the superior performance of our technique in dynamic low-SNR communication environments when compared to existing beam alignment techniques.
    \end{abstract}
    
 
    \section{Introduction}\label{Sec:Intro}

    Millimeter-wave (mmWave) communication has received great attention as a key technology for achieving multi-Gbps data rates by utilizing multi-GHz bandwidths \cite{mmwave_key}. However, a major bottleneck of mmWave communication is the severe free-space path loss and atmospheric absorption, which dramatically decrease the received signal-to-noise ratio (SNR). This limitation hinders the practical use of mmWave communication in a wide range of applications.
    The most widely adopted solution to overcome this bottleneck is to apply directional beamforming using a large antenna array, which provides additional beamforming gain to compensate for the severe attenuations in mmWave channels \cite{mmwave_key,tutorial}.
    To fully leverage the advantage of the directional beamforming, it is crucial to align the beam direction with the mmWave channel direction. 
    This makes beam alignment as a key process to facilitate mmWave communications in modern wireless standards such as 802.11ad and 5G NR \cite{release, tutorial}.  
    A straightforward approach for beam alignment is the exhaustive search which involves transmitting all codewords in a beam codebook. 
    However, due to its reliance on brute-force search, it often results in significant beam training overhead which linearly increases with the codebook size.

    There is a rich literature on efficient beam alignment techniques tailored for mmWave communication systems. 
    Beam alignment techniques can be categorized into two types: (i) hierarchical search, and (ii) environment-aware search. 
    In the hierarchical search, a beam alignment process begins by searching wider beams and gradually narrows down the search based on received reporting information, effectively reducing the overall beam training overhead \cite{hierarchical1,hierarchical2,hierachical_power_alloc}. However, it still needs to search over all potential directions in the environment, thus requiring beam overhead that scales with the size of the antenna array. This limitation is addressed by the environment-aware search, which leverages prior information captured from environments such as user location information \cite{location_beam_FP,location_beam,location_beam_withFP}, sub-6 GHz channel data \cite{sub6_1, sub6_2}, other relevant contextual information \cite{vision,Radar_beam}, and the incorporation of multimodal data \cite{multi_modal_vision_position,multi_modal_magazine}.
    These environment-aware techniques employ supervised learning with sufficient training data to capture and model the complex relationships between side information and probable beam directions \cite{DL_survey}. Then, by leveraging the learned relationships, these techniques reduce the size of the beam search space in order to facilitate fast and efficient beam alignment. Although most of these techniques primarily focus on predicting a single optimal beam, they often select multiple beam candidates (i.e., the top-$k$ method) to improve beam prediction accuracy. In this context, all the aforementioned techniques need to rely on comparing the powers of received signals to determine the optimal beam index. 
    However, this beam determination process exhibits vulnerabilities in low SNR conditions because the received signal with the optimal beam index may not offer the highest power in the presence of noise.
    This limitation is particularly problematic when user equipment (UE) is located at the cell edge or when there is particularly severe path loss due to shadowing effects and blockage \cite{lowSNR_measurements}. Therefore, the development of beam alignment techniques for low-SNR environments is crucial to enhance the cell coverage of mmWave communication systems, which is key to enlarging the applicability of mmWave communications.


    Very limited work has focused on beam alignment for low-SNR mmWave communications \cite{lowSNR_Beam,lowSNR_RACE,lowSNR_RACE_pattern}. 
    In \cite{lowSNR_Beam}, the vulnerability of hierarchical beam search in low SNR was analyzed, emphasizing that the employment of wider beams diminishes beamforming gain, making it challenging to identify the optimal beam index during the early stages in a low SNR regime. 
    This study suggests that exhaustive search with sufficient pilot lengths might be more advantageous in low SNR conditions.
    However, this approach causes a significant beam training overhead, as sufficient pilot lengths must be allocated across all the beam codewords.
    In \cite{lowSNR_RACE,lowSNR_RACE_pattern}, a hierarchical multi-stage structure was introduced with a rate-adaptation method that adaptively allocates more channel measurements in the early stages.
    However, this approach still necessitates scanning all possible directions with multiple rounds, resulting in a large training overhead that hinders its use in fast-fading scenarios with limited channel coherence time.
    Therefore, to fully harness the advantages of mmWave communications even in the low-SNR regime, it is crucial to develop a beam alignment technique that not only provides robustness against noise impact but also reduces the beam training overhead.

    In this paper, we propose a robust beam alignment technique for low-SNR mmWave communications, which reduces beam misalignment probability induced by noise. The key idea of the proposed technique is to repeatedly transmit the most probable beam candidates with the optimized number of beam repetitions, in order to minimize the beam misalignment probability without imposing a significant beam training overhead. To this end, we employ a deep neural network (DNN) that learns the prior probability of the optimal beam at each location. We then analyze the beam misalignment probability by utilizing the beam prior probability estimated by the DNN as the channel prior, guiding the optimal selection of beam candidates. For a special case with two beam candidates, we derive a closed-form expression for the optimal beam repetitions. For a general number of beam candidates, we develop an efficient algorithm to determine the optimal beam allocation among the candidates. Using simulations, we demonstrate the superiority of the proposed beam alignment technique in dynamic low-SNR environments compared to existing techniques.
    The main contributions of this paper are summarized below.
    \begin{itemize}
         \item We introduce a beam repetition strategy for robust beam alignment in low-SNR mmWave communications. In this strategy, we consider the repeated transmission of the most probable beam candidates, while allocating different numbers of beam repetitions for these candidates. We also introduce a DL-based optimization framework for this strategy. In this framework, we utilize a DNN to estimate the prior probability of each beam codeword being optimal at each location and then leverage the estimated prior probabilities as channel priors.

        
        \item We analyze the beam misalignment probability of our beam repetition strategy in low-SNR mmWave communications based on the channel prior information. In this analysis, we characterize the beam misalignment probability as a function of the set of beam candidates and the number of beam repetitions for these candidates. Based on this analysis, we prove that the optimal selection of beam candidates to minimize the beam misalignment probability is to select the beam codewords with the largest beam prior probabilities.
        
        \item We optimize the number of beam repetitions for the selected beam codewords under the constraint of beam training overhead. Specifically, for the special case with two beam candidates, we derive a closed-form expression for the optimal beam allocation among these candidates. For a general number of beam candidates, we develop an efficient algorithm to minimize the upper bound of the beam misalignment probability under certain relaxations. In this development, we demonstrate that finding the optimal beam allocation for this general case involves the use of water-filling algorithms.
    
        \item We discuss practical solutions to overcome challenges that may arise when employing the proposed beam alignment technique, including a feedback strategy and potential solutions for non-stationary channels. Additionally, we explore various possibilities for extending the proposed technique, such as its extension to wideband scenarios and the potential for incorporating additional side information.

        \item  We demonstrate the superiority of the proposed beam alignment technique with comprehensive simulations. We first validate the effectiveness of our optimal beam repetition strategy under the assumption of perfect beam prior probability. We then evaluate the performance improvement achieved by the proposed technique with the DNN-based beam prior probability under dynamic channel conditions with the DeepMIMO dataset \cite{deepmimo}.
        Our technique consistently demonstrates superior performance compared to existing techniques, particularly in dynamic channels and low-SNR communication scenarios, highlighting the effectiveness in practical scenarios.

    \end{itemize}

    
    In this paper, we build upon our previous work \cite{Conference} and extend our analysis by deriving a closed-form solution for optimal beam allocation in a special case with two beam candidates. 
    Furthermore, we demonstrate that the solution for the special case aligns with the general solution we proposed. 
    Additionally, we discuss practical considerations for our proposed beam alignment technique, providing insights to address challenges that may arise in practical scenarios, as well as exploring various possibilities for its extension.
    Finally, we also present additional simulation results to comprehensively validate the proposed technique, highlighting its effectiveness in practical scenarios.


    \section{System Model}\label{Sec:System}
    In this section, we introduce a mmWave communication system considered in our work and then present the challenge of a beam alignment process in low-SNR environments.

    We consider a mmWave multiple-input single-output (MISO) communication system, where the base station (BS) is equipped with a uniform linear array (ULA) consisting of $N$ antenna elements, and the UE has a single-antenna. The BS is located at a fixed position $(\tilde{x}_{\rm BS},\tilde{y}_{\rm BS})$, and performs beam alignment with the UE located at any coordinate $(\tilde{x}_{\ell},\tilde{y}_{\ell})$. In this system, we assume that the BS and UE are synchronized; thereby, the UE is capable of providing feedback to the BS on the transmit beamforming indices during the beam alignment \cite{sync1,sync2}.

    We consider the Saleh-Valenzuela channel model \cite{Channel_model_1,Channel_model_2,Channel_model_3}, in which the mmWave channel between the BS and the UE located at the coordinate $(\tilde{x}_{\ell},\tilde{y}_{\ell})$ is modeled as
    \begin{align}\label{eq:channel_model}
        \mathbf {h}_{\ell} = \sum _{i=1}^{L_{\ell}} \alpha_{ \ell,i} \mathbf {a}(\phi _{\ell,i}),
    \end{align}
    where $L_{\ell}$ is the number of paths, $\alpha_{\ell,i}$ is the complex gain of the $i$-th path, $\phi _{\ell,i}$ is the angle of departure of the $i$-th path, $\mathbf {a}(\phi _{\ell,i})$ is the array response vector of the BS, given by
    \begin{align}\label{eq:array_respons}
    \mathbf {a}(\phi _{\ell,i}) = \frac {1}{\sqrt {N}}\left [{ 1, e^{j\frac {2 \pi }{\lambda }d\sin \phi _{\ell,i} },\ldots, e^{j(N-1)\frac {2      \pi }{\lambda }d\sin \phi _{\ell,i} } }\right]^{\sf T},
    \end{align}
    $\lambda$ is the carrier wavelength, and $d$ is the antenna spacing between elements of the ULA. 
    In this work, we consider a dynamic channel scenario, in which the parameters $L_{\ell}$, $\{\alpha_{\ell,i}\}_{i=1}^{L_{\ell}}$, and $\{\phi_{\ell,i}\}_{i=1}^{L_{\ell}}$ are random variables that follow stationary distributions that can differ across different UE locations.
    For the beam alignment process, we assume that the BS leverages a discrete Fourier transform (DFT) beam codebook, defined as ${\mathbf F} = [{\mathbf f}_1,{\mathbf f}_2,\cdots, {\mathbf f}_N] \in \mathbb{C}^{N \times N}$, which is a widely adopted in 3GPP \cite{DFT}. 
    Each codeword of ${\mathbf F}$ is denoted by
    \begin{align}\label{eq:cordword}
        \mathbf {f}_{c} = \frac {1}{\sqrt {N}} \begin{bmatrix} 1, e^{-j2\pi \cdot \frac{c}{N}}, \cdots, e^{-j 2\pi \cdot \frac{(N-1) \cdot c}{N}} \end{bmatrix}^{\sf T},
    \end{align}
    where $c\in\mathcal{C}=\{1,\ldots,N\}$. Let $\mathcal{T} =\{t_1,\ldots,t_N\}$ be the time index set of the beam training period.

    Consider the beam training process based on the exhaustive search. In this aproach, the BS transmits the beam codeword ${\bf f}_c$ at time slot $t_c\in\mathcal{T}$ for all $c\in\mathcal{C}$. As a result, the baseband received signal
    at the UE at time slot $t_c\in\mathcal{T}$ is given by 
    \begin{align}\label{eq:received_signal}
        y_{\ell}[t_c] = \mathbf{h}_{\ell}^{\sf H}\mathbf {f}_{c}+{z}_{\ell}[t_c], 
    \end{align}
    where ${z}_{\ell}[n]$ is the additive white Gaussian noise with zero mean and variance $\sigma^2$. After receiving the signals $\{y_{\ell}[t_c]\}_{c=1}^N$, the beam index $\hat{c}_{\ell}^{\star}$ with the maximum received power is determined as follows:
    \begin{align}\label{eq:detected_beam}
        \hat{c}_{\ell}^{\star} = \underset{c\in\mathcal{C}}{\arg\!\max} ~|y_{\ell}[t_c]|^2.
    \end{align}
    It should be noted that if the SNR is sufficiently high, the beam index $\hat{c}_{\ell}^{\star}$ determined by the exhaustive search would be the same as the optimal beam index $c_{\ell}^{\star}$, defined as
    \begin{align}\label{eq:optimal_beam}
        c_{\ell}^{\star} = \underset{c\in\mathcal{C}}{\arg\!\max} ~|{\bf h}_{\ell}^{\sf H} {\bf f}_c|^2.
    \end{align}
    Unfortunately, practical mmWave communication systems may operate in a low signal-to-noise ratio (SNR) regime when the UE is located at the cell edge or when there is particularly severe path loss due to shadowing effects and blockage. In such cases, the probability of a beam mismatch between $\hat{c}_{\ell}^{\star}$ and $c_{\ell}^{\star}$ is not negligible. 
    If the beam codeword with the mismatched index is adopted during data transmission, the corresponding data rate can be severely degraded due to a low effective channel gain. 

    \section{Proposed Prior-Aware Robust Beam Alignment Technique}\label{Sec:Robust} 
    In this section, we propose a prior-aware robust beam alignment technique  to reduce a beam mismatch probability in low-SNR mmWave communications.

    \subsection{Beam Repetition Strategy}\label{Sec:Repeat}
    In the proposed technique, we consider a beam repetition strategy which involves repeated transmission of each beam codeword during the beam training period.
    Let $r_{\ell,c}$ be the number of beam repetitions for $c$-th beam codeword ${\bf f}_c$, and let $\mathcal{T}_c = \{t_{c,1},\ldots,t_{c,r_{\ell,c}}\}$ be the set of the time indices allocated for transmitting the $c$-th beam codeword.
    In this scenario, the average of the received signals $\{y_{\ell}[n]\}_{\forall n\in\mathcal{T}_{c}}$ is given by 
    \begin{align}\label{eq:avg_signal}
        \bar{y}_{\ell,c} = \frac{1}{r_{\ell,c}}\sum_{n\in\mathcal{T}_{c}} y_{\ell}[n] = {\bf h}_{\ell}^{\sf H} {\bf f}_c +  \bar{z}_{\ell,c},
    \end{align}
    where $\bar{z}_{\ell,c} = \frac{1}{r_{\ell,c}}\sum_{n\in\mathcal{T}_{c}}z_{\ell}[n]$ is the effective noise that follows the complex Gaussian distribution with zero mean and variance $\sigma^2_{\ell,c}$ (i.e.,  $\bar{z}_{\ell,c} \sim \mathcal{CN}(0,\sigma^2_{\ell,c})$).
    Then the best beam index can be determined using the average received signal $\bar{y}_{\ell,c}$ as follows:
    \begin{align}\label{eq:detected_beam2}
        \hat{c}_{\ell}^{\star} = \underset{c\in\mathcal{C}}{\arg\!\max} ~|\bar{y}_{\ell,c}|^2.
    \end{align}
    It should be noted that the variance of the effective noise $\bar{z}_{\ell,c}$ is given by $\sigma^2_{\ell,c} = {\sigma^2}/{r_{\ell,c}}$, implying that the effective SNR improves as the number of beam repetitions increases. 
    As a result, the beam index $\hat{c}_{\ell}^{\star}$ determined from \eqref{eq:detected_beam2} approaches the optimal beam index $c_{\ell}^{\star}$ as the number $r_{\ell,c}$ of beam repetitions increases for all $c\in\mathcal{C}$, i.e., 
    \begin{align}\label{eq:r_increase}
        \hat{c}_{\ell}^{\star} = \underset{c\in\mathcal{C}}{\arg\!\max} ~|\bar{y}_{\ell,c}|^2 \overset{ \forall r_{\ell,c} \rightarrow \infty}{\longrightarrow} c_{\ell}^{\star}
        = \underset{c\in\mathcal{C}}{\arg\!\max} ~|{\bf h}_{\ell}^{\sf H}{\bf f}_c|^2.
    \end{align}

    The main challenge behind the above beam repetition strategy is that transmitting all the beam codewords with repetition may result in an unaffordable beam training overhead due to limited channel coherence time. 
    This challenge becomes more pronounced in mmWave systems with large antenna arrays, as even a small number of beam repetitions can lead to an excessive beam training overhead.
    To address this challenge, in the following subsections, we will optimize not only the best set of beam candidates, but also the optimal number of beam repetitions for each candidate based on prior knowledge about channel distribution.

    \subsection{Learning the Beam Prior Probability}\label{Sec:DNN} 
    To obtain the channel prior crucial for optimizing the beam repetition strategy in Sec.~\ref{Sec:Repeat}, we leverage a DNN designed to learn the prior probability of the optimal beam across various UE locations.
    Our motivation is that the directionality of a mmWave channel heavily depends on the surrounding environment, and location information can capture pivotal characteristics of this environment. 
    Consequently, location can establish a strong connection with the stochastic characteristics inherent to the mmWave channel. 

    Let ${\bf g}_{\ell} \in[0,1]^N$ be the beam prior probability vector at location $\ell$, defined as 
    \begin{align}
        {\bf g}_{\ell} = \big[ \mathbb{P}(c_{\ell}^\star = 1),\cdots, \mathbb{P}(c_{\ell}^\star = N)\big]^{\sf T},
    \end{align}
    where the $c$-th entry of ${\bf g}_{\ell}$ represents the probability of the $c$-th beam codeword being the optimal at location $\ell$. 
    In the proposed beam alignment technique, a DNN is employed to learn the non-linear mapping relationship from the location $(\tilde{x}_{\ell},\tilde{y}_{\ell})$ to the beam prior probability vector ${\bf g}_{\ell}$, i.e.,
    \begin{align}\label{eq:NN_mapping}
        {\bf{g}}_{\ell} \approx { \boldsymbol {f}}_{ \boldsymbol {\theta }}\big({(\tilde{x}_{\ell},\tilde{y}_{\ell})\big)},
    \end{align}
    where ${ \boldsymbol {f}}_{ \boldsymbol {\theta }}:\mathbb {R}^{2}\longmapsto \mathbb {R}^{N}$ denotes a non-linear mapping process executed by the DNN with the weights $\boldsymbol{\theta}$. For the training of the DNN, it is assumed that sufficient training data samples are attained from beam training history. The training dataset for location $\ell$ is denoted by
    \begin{align}\label{eq:NN_dataset}
    \mathcal{D}_{\ell} = \{((\tilde{x}_{\ell},\tilde{y}_{\ell}),{\bf e}_{{c}^{\star}_{\ell}})\},
    \end{align}
    where ${\bf e}_{{c}^{\star}_{\ell}}$ is a one-hot encoded vector, with the element corresponding to the index of the optimal beam set to one, and the rest set to zero. 
    Then the total training dataset is given by $\mathcal{D} = \cup_{\ell} \mathcal{D}_{\ell}$. 
    Utilizing the training dataset $\mathcal{D}$, the weights $\boldsymbol {\theta}$ of the DNN are trained to minimize the cross-entropy loss function.
    How to utilize the beam prior probability estimated by the DNN for optimizing our beam repetition strategy will be discussed in the sequel.  

    \subsection{Analysis of Beam Misalignment Probability}\label{sec:Analysis_beam_mis}
    We now analyze a beam misalignment probability by utilizing the beam prior probability estimated by our DNN as the channel prior. In this analysis, we aim at characterizing the beam misalignment probability as a function of the beam candidate set and the number of repetitions for each candidate. 
    Let $\mathcal{S}_{\ell} \subset \mathcal{C}$ be the set of the beam candidates for location $\ell$, and let $r_{\ell,\mathcal{S}_{\ell}(i)}$ be the number of beam repetitions allocated for the $i$-th beam candidate in $\mathcal{S}_{\ell}$.
    By the definition, the beam misalignment probability, namely $p_{{\rm miss},\ell}$, can be expressed as the sum of two probabilities:
    \begin{align}\label{eq:prob_divided}
        p_{{\rm miss},\ell} = p_{{\rm miss-sel},\ell} + p_{{\rm miss-det},\ell},
    \end{align}
    where $p_{{\rm miss-sel},\ell}$ denotes the probability of beam miss-selection occurring due to the optimal beam not being included in the beam candidate set, and $p_{{\rm miss-det},\ell}$ denotes the probability of miss-determination occurring due to the failure to determine the optimal beam within the candidate set $\mathcal{S}_{\ell}$ because of the noise.
    These two probabilities can be further characterized as 
    \begin{align}\label{eq:prob_miss_sel_1}
        &p_{{\rm miss-sel},\ell} = \sum_{c\in \mathcal{S}_{\ell}^{c}} \mathbb{P}\big(|{\bf h}_{\ell}^{\sf H}{\bf f}_c|^2 > \max_{j\neq c}|{\bf h}_{\ell}^{\sf H}{\bf f}_j|^2 \big),
    \end{align}
    and
    \begin{align} \label{eq:prob_miss_det_1}
        p_{{\rm miss-det},\ell} 
        = \sum_{c\in \mathcal{S}_{\ell}} \mathbb{P}\Big(& \max_{j\neq c, j\in \mathcal{S}_{\ell}} |\bar{y}_{\ell,j}|^2 > |\bar{y}_{\ell,c}|^2, \nonumber \\ 
        &~~|{\bf h}_{\ell}^{\sf H}{\bf f}_c|^2 > \max_{j\neq c, j\in \mathcal{S}_{\ell}}|{\bf h}_{\ell}^{\sf H}{\bf f}_j|^2\Big), 
    \end{align}
    respectively.

    As can be seen from \eqref{eq:prob_miss_sel_1} and \eqref{eq:prob_miss_det_1}, exact characterization of the beam misalignment probability in \eqref{eq:prob_divided} requires perfect knowledge of the true channel distribution. 
    Unfortunately, in practice, it is very challenging to acquire this knowledge due to diverse communication environments. 
    To circumvent this challenge, we treat the beam prior probability vector $\hat{\mathbf g}_{\ell}$, estimated by our DNN, as if it is a true beam probability vector.
    Then, one can easily see that the beam prior probability vector $\hat{\mathbf g}_{\ell}$ holds when the channel distribution at location $\ell$ is given by
    \begin{align}\label{eq:channel_assumption}
        \mathbb{P}({\bf h}_{\ell} =  \alpha_{\ell} {\bf f}_c) =\hat{g}_{\ell,c},~\forall c\in\{1,\ldots,N\}.
    \end{align}
    Motivated by the above fact, we employ the channel model in \eqref{eq:channel_assumption} for characterizing the beam miss-selection and miss-determination probabilities. 
    From \eqref{eq:channel_assumption}, the beam miss-selection probability can be characterized as 
    \begin{align}\label{eq:prob_miss_sel_2}
        &p_{{\rm miss-sel},\ell} \nonumber \\
        &= \sum_{c\in \mathcal{S}_{\ell}^c} \sum_{k=1}^N \mathbb{P}\big(|{\bf h}_{\ell}^{\sf H}{\bf f}_c|^2 > \max_{j\neq c}|{\bf h}_{\ell}^{\sf H}{\bf f}_j|^2 \big| {\bf h}_{\ell} {=} \alpha_{\ell}{\bf f}_k \big) \mathbb{P}({\bf h}_{\ell} {=} \alpha_{\ell}{\bf f}_k ) \nonumber\\
        &= \sum_{c\in \mathcal{S}_{\ell}^c}  \hat{g}_{\ell,c} = 1 - \sum_{c\in \mathcal{S}_{\ell}}  \hat{g}_{\ell,c}.
    \end{align} 
    Similarly, the beam miss-determination probability in \eqref{eq:prob_miss_det_1} can be rewritten as
    \begin{align}\label{eq:prob_miss_det_2}
        &p_{{\rm miss-det},\ell}  \nonumber \\
        &= \sum_{c\in \mathcal{S}_{\ell}} \sum_{k=1}^N \mathbb{P}\bigg(\max_{j\neq c, j\in \mathcal{S}_{\ell}} |\bar{y}_{\ell,j}|^2 > |\bar{y}_{\ell,c}|^2, \nonumber\\
        &\qquad |{\bf h}_{\ell}^{\sf H}{\bf f}_c|^2 > \max_{j\neq c, j\in \mathcal{S}_{\ell}}|{\bf h}_{\ell}^{\sf H}{\bf f}_j|^2 \bigg| {\bf h}_{\ell} = \alpha_{\ell}{\bf f}_k \bigg) \mathbb{P}({\bf h}_{\ell} = \alpha_{\ell}{\bf f}_k ) \nonumber \\
        &= \sum_{c\in \mathcal{S}_{\ell}}  \mathbb{P}\Big(\max_{j\neq c, j\in \mathcal{S}} |\alpha_{\ell}{\bf f}_c^{\sf H}{\bf f}_j + \bar{z}_{\ell,j}|^2 > |\alpha_{\ell}{\bf f}_c^{\sf H}{\bf f}_c + \bar{z}_{\ell,c}|^2\Big)\hat{g}_{\ell,c}.
    \end{align}  
    Recall that we adopt the DFT codebook with a size equal to the number of transmit antennas.
    By the property of the DFT codebook, all codewords are orthogonal, satisfying
    ${\bf{f}}_{c}^{\sf H}{\bf{f}}_{j}=0$, $\forall c \neq j$. 
    Using this, the upper bound of the beam miss-determination probability can be derived as 
    \begin{align}\label{eq:prob_miss_det_3}
        &p_{{\rm miss-det},\ell} 
        =  \sum_{c\in \mathcal{S}_{\ell}} \mathbb{P}\Big(\max_{j\neq c, j\in \mathcal{S}_{\ell}} |\bar{z}_{\ell,j}|^2 > |\alpha_{\ell} + \bar{z}_{\ell,c}|^2\Big)\hat{g}_{\ell,c}  \nonumber \\
        &\overset{(a)}{\leq} \sum_{c\in \mathcal{S}_{\ell}}  \hat{g}_{\ell,c}  \sum_{j\neq c, j\in \mathcal{S}_{\ell}} \mathbb{P}\big( |\bar{z}_{\ell,j}|^2 > |\alpha_{\ell} + \bar{z}_{\ell,c}|^2\big),
    \end{align} 
    where $(a)$ follows from the union bound. 
    Utilizing the fact that $\bar{z}_{\ell,c} \sim \mathcal{CN}(0,\sigma^{2}/r_{\ell,c})$, we characterize the pair-wise miss-determination probability $\mathbb{P}\big( |\bar{z}_{\ell,j}|^2 > |\alpha_{\ell} + \bar{z}_{\ell,c}|^2\big)$ in \eqref{eq:prob_miss_det_3}. This result is given in the following lemma:
    \begin{lem}\label{lem:pair_wise_prob}
    The pair-wise miss-determination probability $\mathbb{P}\big( |\bar{z}_{\ell,j}|^2 > |\alpha_{\ell} + \bar{z}_{\ell,c}|^2\big)$ is computed as
    \begin{align}\label{eq:prop_1}
        \mathbb{P}\big( |\bar{z}_{\ell,j}|^2 > &|\alpha_{\ell} + \bar{z}_{\ell,c}|^2\big)  =\frac{r_{\ell,c}}{r_{\ell,c}+r_{\ell,j}}{\rm exp}\bigg({-\frac{r_{\ell,c}r_{\ell,j}}{r_{\ell,c}+r_{\ell,j}}\rho_{\ell}}\bigg),
    \end{align}
    where $\rho_{\ell}=|\alpha_{\ell}|^2/\sigma^2$ is the received SNR at location $\ell$. 
    \end{lem}
    \begin{IEEEproof}
        See Appendix~\ref{apdx:pair_wise_prob}.
    \end{IEEEproof}
    \vspace{1mm}
    
    Applying {\bf Lemma \ref{lem:pair_wise_prob}} into \eqref{eq:prob_miss_det_3}, the upper bound of the beam miss-determination probability is obtained as
    \begin{align}\label{eq:prob_mis_det}
        p_{{\rm miss-det},\ell} 
        \leq  \underbrace{\sum_{c\in \mathcal{S}} \hat{g}_{\ell,c} \sum_{j\neq c, j\in \mathcal{S}}\kappa_{\ell,c,j}{\rm exp}\Big(-{\kappa_{\ell,c,j}r_{\ell,j}\rho_{\ell}}\Big)}_{\triangleq \hat{p}_{{\rm miss-det},\ell}},
    \end{align} 
    where $\kappa_{\ell,c,j}=r_{\ell,c}/(r_{\ell,c}+r_{\ell,j})$. Consequently, the upper bound of the beam misalignment probability is derived as
    \begin{align}\label{eq:upper_overall}
        &p_{{\rm miss},\ell}= p_{{\rm miss-sel},\ell} + p_{{\rm miss-det},\ell}  \nonumber \\
        &\!\leq \underbrace{{1 - \sum_{c\in \mathcal{S}_{\ell}}  \hat{g}_{\ell,c} \bigg( 1 - \sum_{j\neq c, j\in \mathcal{S}}
        \kappa_{\ell,c,j}{\rm exp}\Big(-{\kappa_{\ell,c,j}r_{\ell,j}\rho_{\ell}}\Big)\bigg) }}_{\triangleq \hat{p}_{{\rm miss},\ell}}.
    \end{align}
    As can be seen in \eqref{eq:upper_overall}, there exists a trade-off between the size of the beam candidate set and the number of beam repetitions for each candidate. 
    For example, if we increase the size of the beam candidate set, this decreases the beam miss-selection probability as the likelihood of including the optimal beam in the candidate set increases.
    However, for a given beam training overhead, increasing the size of the beam candidate set naturally decreases the number of beam repetitions for each candidate. This leads to the increase in the beam miss-determination probability because the less number of beam repetitions decreases the SNR of the received signal in \eqref{eq:avg_signal}.
    Therefore, it is crucial to optimize both the selection of beam candidates and the number of beam repetitions for each candidate, in order to minimize the beam misalignment probability. 
    
    \subsection{Optimization Problem Formulation}\label{Sec:Formulation}
    Based on the analysis of the beam misalignment probability in Sec. \ref{sec:Analysis_beam_mis}, we optimize both the selection of beam candidates and the number of repetitions for each candidate, to minimize the upper bound in \eqref{eq:upper_overall}.
    From \eqref{eq:upper_overall}, one can easily notice that the upper bound $\hat{p}_{{\rm miss},\ell}$ is minimized when the beam codewords with the highest prior probabilities are chosen as beam candidates. 
    Therefore, for a fixed size $S$, the optimal set of the beam candidates is determined as
    \begin{align}\label{eq:candidate_set}
        \mathcal{S}_{\ell}^\star (S) = \{c_{\ell,1}^\star,c_{\ell,2}^\star,\ldots,c_{\ell,S}^\star\},
    \end{align}
    where $c_{\ell,i}^\star$ is the index of the beam codeword with the $i$-th largest $\hat{g}_{\ell,i}$, and $S$ is the size of the beam candidate set. 

    Given the optimal set in \eqref{eq:candidate_set}, the remaining task is to optimize the size $S$ of the beam candidate set and the number of beam repetitions for each candidate in $\mathcal{S}_{\ell}^\star (S)$. 
    Define ${\bf r}_{\ell}(S) = [r_{\ell, c_{\ell,1}^\star},\cdots, r_{\ell, c_{\ell,S}^\star}]^{\sf T}$ as a {\em beam allocation} vector which represents the allocation of the beam repetitions for the beam candidates in $\mathcal{S}_{\ell}^\star(S)$. 
    Then, the optimization problem to determine $S$ and ${\bf r}_{\ell}(S)$ to minimize the beam misalignment probability for a given beam training overhead $R_{\rm sum}$ is formulated as
    \begin{align}\label{eq:p_hat_opt}
        \underset{{S},{\bf r}_{\ell}(S)}{\arg\!\min}~\hat{p}_{{\rm miss},\ell}, 
        ~~\text{s.t.}~~{\bf 1}^{\sf T}{\bf r}_{\ell}(S) \leq R_{\rm sum}.
    \end{align}

    A straightforward way to solve the optimization problem in \eqref{eq:p_hat_opt} is to take an exhaustive searching approach which necessitates comparing all possible choices of $S$ and ${r}_{\ell}(S)$. This approach, however, involves tremendous computational complexity because even for a fixed size $S$, the number of partitions of $R_{\rm sum}$ into $S$ positive integers exceeds $\frac{1}{S !}\left(\frac{R_{\rm sum}-1}{S}\right)^S $ \cite{numtheory}.  
    To address this problem, in what follows, we first derive the optimal beam allocation for a special case with $S=2$ and then develop a computationally-efficient algorithm to determine the near-optimal beam allocation for the general case of $S>2$. 

    \subsection{Closed-Form Solution for Optimal Beam Allocation with $S=2$}\label{Sec:Optimization}
    In this subsection, we characterize the closed-form expression for the optimal beam allocation to minimize the beam misalignment probability when $S=2$. 
    Without loss of generality, suppose that the optimal candidate set is given by $\mathcal{S}_{\ell}^\star (2) = \{c_{\ell,1}^\star, c_{\ell,2}^\star\} = \{1,2\}$. 
    Then the optimization problem in \eqref{eq:p_hat_opt} for $S=2$ can be reformulated as
    \begin{align}\label{eq:p_hat_opt_S_2}
        \underset{r_{\ell, 1}, r_{\ell, 2}}{\arg\!\min}~&\hat{p}_{{\rm miss-det},\ell} 
        \nonumber 
        \\
        \text{s.t.}~~~ &r_{\ell, 1}+ r_{\ell, 2} \leq R_{\rm sum}.
    \end{align}
    For $S=2$, the beam miss-determination probability $\hat{p}_{{\rm miss-det},\ell}$ is rewritten as
    \begin{align}\label{eq:p_miss_det_S_2}
        \hat{p}_{{\rm miss-det},\ell} = \frac{r_{\ell,1} \hat{g}_{\ell,1}+r_{\ell,2} \hat{g}_{\ell,2}}{R_{\rm sum}}{\rm exp}\left(-{\frac{r_{\ell,1}r_{\ell,2}}{R_{\rm sum}}\rho_{\ell}}\right).
    \end{align}
    Due to the integer constraints on $r_{\ell, 1}$ and $r_{\ell, 2}$, it is difficult to characterize the closed-form solution of the problem in \eqref{eq:p_hat_opt_S_2}. 
    To circumvent this difficulty, we relax these constraints by treating $r_{\ell,1}$ and $r_{\ell,2}$ as non-negative real numbers. 
    Define $x_{\ell} \triangleq {r}_{\ell,2}/{r}_{\ell,1}$ as the beam-allocation ratio between two candidates, where $x_{\ell}$ is a non-negative real number. 
    Given beam training overhead $R_{\rm sum}$, the beam allocations can be expressed as ${r}_{\ell,1} = R_{\rm sum}/(x_{\ell}+1), \ {r}_{\ell,2} = x_{\ell}R_{\rm sum}/(x_{\ell}+1)$.
    Utilizing these expressions, the beam miss-determination probability in \eqref{eq:p_miss_det_S_2} is rewritten as a function of $x_{\ell}$:
    \begin{align}\label{eq:p_miss_det_S_2_relaxation}
    &\hat{p}_{{\rm miss-det},\ell} (x_{\ell})\nonumber\\
    &=\left(\frac{\hat{g}_{\ell,1}R_{\rm sum}+\hat{g}_{\ell,2}{x_{\ell}R_{\rm sum}}}{(x_{\ell}+1)R_{\rm sum}}\right){\rm exp}{\left({-\frac{x_{\ell} R_{\rm sum}}{(x_{\ell}+1)^2}\rho_{\ell}}\right)}
    \nonumber \\&= \frac{(k_{\ell}+x_{\ell})\hat{g}_{\ell,2}}{x_{\ell}+1}{\rm exp}{\left({-\frac{x_{\ell}}{(x_{\ell}+1)^2}\beta_{\ell}}\right)},
    \end{align}
    where $\beta_{\ell} = R_{\rm sum}\rho_{\ell}$ and $k_{\ell} = \hat{g}_{\ell,1}/\hat{g}_{\ell,2}$. 
    The optimal beam-allocation ratio, namely $x^{\star}_{\ell}$, that minimizes $\hat{p}_{{\rm miss-det},\ell}(x)$ is characterized as a closed-form expression, as given in the following theorem:
    \vspace{1mm}
    \begin{thm}\label{thm:x_opt_S_2}
    If $\hat{g}_{\ell,1} > \hat{g}_{\ell,2}$, the optimal beam-allocation ratio $x^{\star}_{\ell}$ 
    satisfies $x^{\star}_{\ell}>1$, and is determined as 
    \begin{align}
        x^{\star}_{\ell} =
        \begin{cases}
        \frac{-a_{\ell,1}+\sqrt{a_{\ell,1}^2-4a_{\ell,2}a_{\ell,0}}}{2a_{\ell,2}}, &\text{if } \beta_{\ell}>k_{\ell}-1, \\
        -\frac{a_{\ell,0}}{a_{\ell,1}},  &\text{if }  \beta_{\ell}=k_{\ell}-1, \  \beta_{\ell}>2, \\
        \frac{-a_{\ell,1}-\sqrt{a_{\ell,1}^2-4a_{\ell,2}a_{\ell,0}}}{2a_{\ell,2}}, &\text{if } 2<\beta_{\ell}<k_{\ell}-1, \ D>0,\\
        \infty, & \text{otherwise},
        \end{cases}
    \end{align}
    where $a_{\ell,2}=\beta_{\ell}-k_{\ell}+1$, $a_{\ell,1}=(\beta_{\ell}-2)(k_{\ell}-1)$, $a_{\ell,0}=-(\beta_{\ell}+1)k_{\ell}+1$, and $D = a_{\ell,1}^2-4a_{\ell,2}a_{\ell,0}$.
    \end{thm} 
    \begin{IEEEproof}
    See Appendix~\ref{apdx:x_opt_S_2}.
    \end{IEEEproof}
    \vspace{1mm}
    Subsequently, since we have relaxed the constraints into the real domain, the original constraints can be re-applied to find the optimal values as follows:
    \begin{align}\label{eq:r_S_2}
        (r_{\ell,1}^\star, r_{\ell,2}^\star) =
        \begin{cases}
         \left(\bigg\lfloor \dfrac{R_{\rm sum}}{x^{\star}_{\ell}+1} \bigg\rfloor, \bigg\lfloor \dfrac{x^{\star}_{\ell}R_{\rm sum}}{x^{\star}_{\ell}+1} \bigg\rfloor\right),
         &\text{if } \dfrac{R_{\rm sum}}{x^{\star}_{\ell}+1}>1, \\
         \\
        (1, R_{\rm sum}-1),  &\text{if }  \dfrac{R_{\rm sum}}{x^{\star}_{\ell}+1} \leq 1.  
        \end{cases}
    \end{align}
    It is worth noting that if ${R_{\rm sum}}/({x^{\star}_{\ell}+1})>1$, the floor function may result in a remainder of 1 for the training overhead. This can be compensated by adding 1 to the side that makes the ratio $r_{\ell,2}^\star/r_{\ell,1}^\star$ closer to $x^{\star}_{\ell}$.

    Our analysis in {\bf Theorem 1} also reveals the important property of the optimal beam allocation, which is described in the following corollary:
    \vspace{1mm}
    \begin{cor}\label{cor:r_opt_S_2}
    If $\hat{g}_{\ell,1} > \hat{g}_{\ell,2}$, the solution $(r_{\ell,1}^\star, r_{\ell,2}^\star)$ of \eqref{eq:p_hat_opt_S_2}  satisfies $r_{\ell, 1}^\star \leq r_{\ell, 2}^\star$.
    \end{cor}
    \begin{IEEEproof}
    The optimal ratio satisfies ${x^{\star}_{\ell}}>1$ from {\bf Theorem \ref{thm:x_opt_S_2}}. Consequently, the optimal allocation $(r_{\ell,1}^\star, r_{\ell,2}^\star)$ in \eqref{eq:r_S_2} satisfy $r_{\ell, 1}^\star \leq r_{\ell, 2}^\star$.    
    \end{IEEEproof}
    \vspace{1mm}
    
    {\bf Corollary 1} implies that when beam candidates have different beam-prior probabilities, non-uniform beam allocation across the candidates is necessary to minimize the beam misalignment probability. It is noteworthy that the directionality of a mmWave channel typically results in varying beam-prior probabilities at different locations. Therefore, our analysis justifies the need for optimizing beam allocation among the candidates when employing the beam repetition strategy.

    \subsection{Efficient Algorithm for Optimal Beam Allocation with $S>2$}\label{Sec:Optimization_general}
    In this subsection, we put forth a computationally-efficient algorithm to determine the near-optimal beam allocation for the general case with multiple beam candidates (i.e., $S>2$). For the purpose of analytical tractability, we start by introducing an approximate upper bound for $\hat{p}_{{\rm miss-det},\ell}$ as follows:
    \begin{align}\label{eq:prob_mis_det_general}
        &\hat{p}_{{\rm miss-det},\ell} \nonumber\\
        &\qquad =  {\sum_{c\in \mathcal{S}} \hat{g}_{\ell,c} \sum_{j\neq c, j\in \mathcal{S}}\kappa_{\ell,c,j}{\rm exp}\big(-{\kappa_{\ell,c,j}r_{\ell,j}\rho_{\ell}}\big)}
        \nonumber\\
        &\qquad \overset{(a)} \lesssim \sum_{c\in \mathcal{S}} \hat{g}_{\ell,c} \sum_{j\neq c, j\in \mathcal{S}}{\rm exp}\big(-{r_{\ell,j}}{\rho_{\ell}}\big)
        \nonumber \\
        &\qquad = \sum_{j\in \mathcal{S}_{\ell}} \exp\left(-r_{\ell,j}{\rho_{\ell}}\right) \sum_{c\neq j, c\in \mathcal{S}_{\ell}} \hat{g}_{\ell,c},
    \end{align} 
    where the approximate upper bound $(a)$ becomes exact in a low-SNR regime with limited beam training overhead, specifically when the condition $r_{\ell,j}\rho_{\ell} \leq 1$ holds. This is because a function $x e^{-k x}$ is monotonically increasing over the interval $[0,1/k]$ for $k>0$, while $\kappa_{\ell,c,j}$ lies within this interval when $r_{\ell,j}\rho_{\ell} \leq 1$.  
    Under the above bound, for a given set size $S$, the optimal allocation for the beam repetitions is obtained by solving the following problem:
    \begin{align}\label{eq:omega_opt}
        \underset{{\bf r}_{\ell}(S)}{\arg\!\min}~&\sum_{c\in \mathcal{S}_\ell^\star (S)} \omega_{\ell,c} \exp\left(-r_{\ell,c}{\rho_{\ell}}\right),  \nonumber \\
        \text{s.t.}~~~ &{\bf 1}^{\sf T}{\bf r}_{\ell}(S) \leq R_{\rm sum}, 
    \end{align}
    where $\omega_{\ell,c} = \sum_{j\in \mathcal{S}_\ell^\star (S)} \hat{g}_{\ell,j} -\hat{g}_{\ell,c} \in [0,1]$.
    The above problem is still an NP-hard problem as it remains non-convex with integer variables.  
    To circumvent this challenge, we consider a further upper bound of the objective function in \eqref{eq:omega_opt} as follows:
    \begin{align}\label{eq:exp_opt}
        \sum_{c\in \mathcal{S}_{\ell}^\star (S)} \omega_{\ell,c} \exp\left(-r_{\ell,c}{\rho_{\ell}}\right)
        &=\sum_{c\in \mathcal{S}_{\ell}^\star (S)}\psi_{\ell}^{N_{\ell,c}  + r_{\ell,c}} \nonumber \\
        &\leq S \psi_{\ell}^{ \min_{c\in \mathcal{S}_{\ell}^\star (S)}\{N_{\ell,c} + r_{\ell,c}\}},
    \end{align}
    where $\psi_{\ell} = \exp(-\rho_{\ell})$ and $N_{\ell,c}= \frac{\log w_{\ell,c}}{\log \psi_{\ell}}$.
    Then, instead of solving \eqref{eq:omega_opt}, we minimize the upper bound of the objective function in \eqref{eq:omega_opt} by solving the following problem:
    \begin{align}\label{eq:final_opt}
         ({\bf P}) ~\underset{{\bf r}_{\ell}(S)}{\arg\!\max}~&\min_{c\in \mathcal{S}_{\ell}^\star (S)} ~\{{N_{\ell,c} + r_{\ell,c}}\}, \nonumber \\
        \text{s.t.}~~~~ &{\bf 1}^{\sf T}{\bf r}_{\ell}(S) \leq R_{\rm sum}. 
    \end{align}
    Since the problem $({\bf P})$ is the water-filling problem \cite{water_fast,water_bisection}, where the value of ${r_{\ell,c}}$ is restricted to an integer value, we can readily solve this problem by leveraging the Lagrangian method.
    Then, the optimal number of the beam repetitions for the candidate $c\in\mathcal{S}_{\ell}^\star (S)$ is obtained as   
    \begin{align}\label{eq:opt_sol}
        r_{\ell,c}  = \lfloor\nu^\star - N_{\ell,c}\rfloor + 1, 
    \end{align}
    where $\nu^\star$ is the optimal Lagrange multiplier satisfying
    \begin{align}\label{eq:opt_sol_const}
        \sum_{c\in\mathcal{S}_{\ell}^\star (S)} \nu^\star - N_{\ell,c} = R_{\rm sum} - S.
    \end{align}
    Various water-filling algorithms, such as the fast water-filling algorithm \cite{water_fast} and the bisection search algorithm \cite{water_bisection}, can be used to obtain the optimal Lagrange multiplier $\nu^\star$. It is worth noting that the floor function used in \eqref{eq:opt_sol} may result in remainders of the training overhead, i.e., $R_{\rm sum} - {\bf 1}^{\sf T}{\bf r}_{\ell}(S) $. A simple solution to compensate for this is to distribute the remainder with each $r_{\ell,c}$ based on the difference $d_{\ell,c} = r_{\ell,c} - (\nu^\star - N_{\ell,c})$ \cite{round}. Then the overhead constraints can be met while maximizing $\min_{c\in \mathcal{S}_{\ell}^\star (S)} ~\{{N_{\ell,c} + r_{\ell,c}}\}$.
    
    The general solution exhibits a similar trend to the analysis for the case with $S=2$ presented in Appendix \ref{apdx:x_opt_S_2}.
    Specifically, in the scenario where the SNR and beam training overhead are high, the values $\{N_{\ell,c}\}$ become moderate, and resources tend to be more evenly distributed across the candidates, causing $r_{\ell,c} \approx r_{\ell,j}$, $\forall c \neq j$.
    In contrast, when the difference in prior probabilities becomes more pronounced, the differences among the values $\{N_{\ell,c}\}$ become large, and the water-filling solution allocates more resources to the beam candidate with a lower prior probability, leading to $r_{\ell,c} \gg r_{\ell,j}$ with $g_{\ell,c} \ll g_{\ell,j}$, $\forall c \neq j$.

    The objective function $\hat{p}_{{\rm miss},\ell}$ in \eqref{eq:upper_overall} can be evaluated by utilizing the optimal beam allocation ${\bf r}_{\ell}(S)$ for a fixed set size $S$. Then, we choose the best candidate set $\mathcal{S}_{\ell}^\star$ and the corresponding beam allocation ${\bf r}_{\ell}^{\star}$ by repeating this procedure for increasing set sizes and selecting the set that yields the smallest value of $\hat{p}_{{\rm miss},\ell}$.
    The overall optimization algorithm is summarized in {\bf Algorithm~\ref{alg:overall}}.
    \begin{algorithm}[t]
        \caption{Optimal Selection of Beam Candidate Set and Beam Allocation Vector}\label{alg:overall}
    	{\small
    	{\begin{algorithmic}[1]
            \REQUIRE {Beam probability} $\hat{\bf g}_{\ell}$, total overhead $R_{\rm{sum}}$
            \ENSURE Beam candidate set $\mathcal{S}^{\star}_{\ell}$, beam allocation vector ${\bf r}_{\ell}^{\star}$
            \STATE $k_1 = \argmax_{j\in \mathcal{C}}\hat{g}_{\ell,j} $
            \STATE $\mathcal{S}_{\ell}^{\star}(1) = \{k_1\}$; $r_{\ell,k_1}=R_{\rm sum}$
            \STATE  $\hat{p}_{{\rm miss},\ell}{(1)} = 1-\hat{g}_{\ell,k_1}$
            \FOR {$S=2$ to $N$}
            \STATE $k_S = \argmax_{j \in \mathcal{C}\setminus \mathcal{S}^{\star}_{\ell}(S-1)} \hat{g}_{\ell,j}$
            \STATE $\mathcal{S}_{\ell}^{\star}(S) =\mathcal{S}_{\ell}^{\star}(S-1) \cup \{k_S\}$
            \STATE Determine ${\bf{r}}_{\ell}{(S)}$ by solving the problem $({\bf P})$ 
            \STATE Compute $\hat{p}_{{\rm miss},\ell}{(S)}$ from \eqref{eq:upper_overall}
            \STATE {\bf if} {$\hat{p}_{{\rm miss},\ell}{(S-1)} <\hat{p}_{{\rm miss},\ell}{(S)}$} {\bf then}
            \STATE ~~$\mathcal{S}^{\star}_{\ell} = \mathcal{S}_{\ell}^{\star}{(S-1)}$; ${\bf r}^{\star}_{\ell} = {\bf r}_{\ell}{(S-1)}$
            \STATE ~~Break the loop
            \STATE {\bf else}
            \STATE ~~$\mathcal{S}^{\star}_{\ell} = \mathcal{S}_{\ell}^{\star}{(S)}$; ${\bf r}^{\star}_{\ell} = {\bf r}_{\ell}{(S)}$
            \STATE {\bf end}
            \ENDFOR
    	\end{algorithmic}}}
    \end{algorithm}
    
    \subsection{Summary}\label{Sec:Summary} 
    \begin{figure}[t]
        \centering
        {\epsfig{file=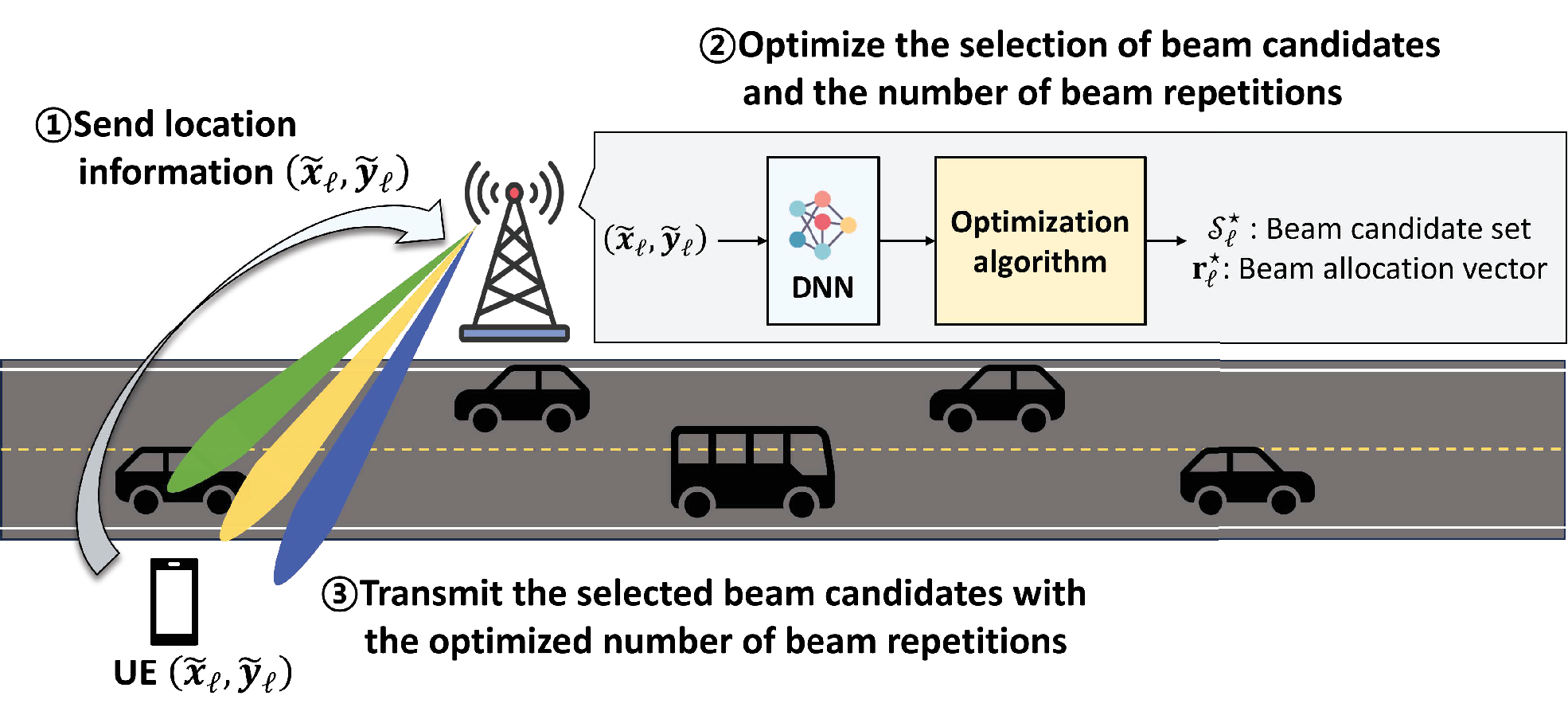,width=9cm}}
        \caption{An illustration of the overall process of the proposed beam alignment technique.} 
        \label{fig:system_model}\vspace{-3mm}
    \end{figure}
    This subsection describes the overall process of the proposed beam alignment technique which consists of the following four steps.
    Firstly, the UE sends its location information to the BS.
    Secondly, the BS utilizes the trained DNN model to obtain the beam prior probability vector $\hat{\bf g}_{\ell}$ by inputting $(\tilde{x}_{\ell},\tilde{y}_{\ell})$. 
    Thirdly, using {\bf Algorithm~\ref{alg:overall}}, the BS solves an optimization problem to determine the set of beam candidates and the beam allocation vector. 
    It then transmits the corresponding codewords with repetition.
    Finally, the UE determines the optimal beam index from \eqref{eq:detected_beam2} and sends the selected optimal index to the BS, thereby completing the beam alignment process.
    The overall process is illustrated in Fig. \ref{fig:system_model}.
    
    \section{Practical Considerations}\label{Sec:Practical} 
    In this section, we discuss some practical considerations for the proposed beam alignment technique. 
    These considerations include approaches to overcome challenges that may arise in practical scenarios, as well as explore the various possibilities for extending the proposed technique.
    
    \subsection{Feedback Strategy for Beam Repetition Information}\label{Sec:Feedback} 
    For the proposed technique to be applied in practical scenarios, the BS needs to convey the information about the optimal set size $S$ and the optimized beam allocation vector ${\bf r}^{\star}_{\ell}$ to the UE.
    Assume that the optimal set size is determined as $S$. Then the information to be conveyed is represented as
    \begin{align}\label{eq:Feedback_information}
        &{\bf r}_{\ell}^\star = \{r_{c_{\ell,1}^\star}^{\star},r_{c_{\ell,2}^\star}^\star,\ldots,r_{c_{\ell,S}^\star}^{\star}\},
    \end{align}
    where $c_{\ell,i}^\star$ is the index of the beam codeword with the $i$-th largest $\hat{g}_{\ell,i}$.
    However, transmitting repetition numbers for every beam candidate also imposes additional communication overhead. 
    To mitigate this overhead, we suggest to compress the information of the beam allocation vector using a linear regression method. 
    As mentioned in Sec.~\ref{Sec:Optimization_general}, since the water-filling solution allocates more repetitions to beams with lower prior probabilities, the entries of the beam allocation vector in \eqref{eq:Feedback_information} follow in ascending order.  
    Utilizing this fact, one cane adopt a linear regression technique to compress the beam allocation vector. The associated linear regression parameters can be obtained by solving the following problem:
    \begin{align}\label{eq:regression_parameters}
        \underset{ \hat{\beta}_0, \hat{\beta}_1}{\arg\!\min}~\sum_{j\in\{1,\ldots,S\}} ({r}_{c_{\ell,j}^\star}^{\star} - \hat{\beta}_0 - \hat{\beta}_1 j)^2.
    \end{align}
    To obtain these parameters, the least squares method can be employed.
    Then, the approximate repetition number for each beam is given by
    \begin{align}
        \hat{r}_{c_{\ell,j}^\star}^{\star} = \lfloor{\hat{\beta}_0 + \hat{\beta}_1 j}\rfloor, ~\forall j\in\{1,\ldots,S\}.
    \end{align}
    Instead of sending the information for the entire beam repetition numbers, transmitting only the curve-fitting parameters $\hat{\beta}_0$, $\hat{\beta}_1$, and the optimal set size $S$ allows the UE to estimate the repetition number for each beam. For consistency, the BS also needs to carry out the beam alignment process according to the approximate repetition number. 
    
    
    \subsection{Uniform Planar Array Configurations}
    While we have focused on the ULA configuration, it is worth noting that the proposed beam alignment technique can also be applied to a uniform planar array (UPA) configuration \cite{location_beam_FP}. 
    Such configuration facilitates beamforming capabilities in both the horizontal (azimuth) and vertical (elevation) planes, further enhancing system performance. 
    For a UPA setup with dimensions $N = N_\mathrm{x} \times N_\mathrm{y}$, the Saleh-Valenzuela channel model can be expressed as follows:
    \begin{align}
        \mathbf {h}_{\ell} = \sum _{i=1}^{L_{\ell}} \alpha_{ \ell,i} \mathbf {a} (\theta _{\ell,i},\phi _{\ell,i}),
     \end{align}
    where $\theta_{\ell,i}$ and $\phi_{\ell,i}$ are the elevation and azimuth angles of departure for the $i$-th path, and the array response vector $\mathbf {a}(\theta _{\ell,i},\phi _{\ell,i})$ is given by
    \begin{align}
    \mathbf {a}(\theta _{\ell,i},\phi _{\ell,i}) = \frac{1}{\sqrt{N_\mathrm{x}N_\mathrm{y}}} \begin{bmatrix}1 \\e^{\mathrm{j}\Omega _\mathrm{y}} \\\vdots \\e^{\mathrm{j}(N_\mathrm{y}-1)\Omega _\mathrm{y}}\end{bmatrix} \otimes \begin{bmatrix}1 \\e^{\mathrm{j}\Omega _\mathrm{x}} \\\vdots \\e^{\mathrm{j}(N_\mathrm{x}-1)\Omega _\mathrm{x}}\end{bmatrix}, 
    \end{align}
    where $\otimes$ represents the Kronecker product and the orientation parameters for the horizontal (azimuth) and vertical (elevation) planes are given by
    \begin{align}
    \Omega\mathrm{y} = \frac{2\pi d \sin(\theta_{\ell,i}) \sin(\phi_{\ell,i})}{\lambda}, \
    \Omega\mathrm{x} = \frac{2\pi d \sin(\theta_{\ell,i}) \cos(\phi_{\ell,i})}{\lambda},
    \end{align}
    respectively.
    For the UPA setup, we can adopt the two-dimensional (2D) DFT codebook ${\mathbf F} = [{\mathbf f}_1,{\mathbf f}_2,\cdots, {\mathbf f}_N]$ where the $c$-th codeword is given by $\mathbf {f}_{c} = \left [{\mathbf {F}_{\mathrm {x}}\otimes \mathbf {F}_{\mathrm {y}}}\right]_{:, c}$,
    where ${\mathbf {F}_{\mathrm {x}}}$ and ${\mathbf {F}_{\mathrm {y}}}$ are the $N_\mathrm{x} \times N_\mathrm{x}$ DFT and $N_\mathrm{y} \times N_\mathrm{y}$ DFT codebooks, respectively. 
    For the UPA setup, the proposed technique in Sec.~\ref{Sec:Robust} can be applied in the same manner as in the case of the ULA setup.

    \subsection{Extension to Wideband Scenarios}
    MmWave communication systems often employ large signal bandwidths, resulting in wideband (i.e., frequency-selective) channels. By adopting an orthogonal frequency division multiplexing (OFDM) waveform, these wideband channels can be divided into ${K}$ parallel narrowband subchannels, where ${K}$ is the number of OFDM subcarriers. Since analog beamforming is implemented in the analog RF domain, the same beamforming is applied across all subchannels. Therefore, for mmWave OFDM systems, the optimal beam index can be determined by comparing the average power of the received signals measured across all subchannels \cite{Wideband_OFDM}, i.e.,
    \begin{align}
        \hat{c}_{\ell}^{\star} = \underset{c\in\mathcal{C}}{\arg\!\max} ~\frac{1}{{K}}\sum_{{k}=1}^{{{K}}}|\bar{y}_{\ell,c}[{k}]|^2,
    \end{align}
    where $\bar{y}_{\ell,c}[{k}]$ represents an average received signal for the $c$-th beam codeword at the ${k}$-th subchannel, given by
    \begin{align} 
        \bar{y}_{\ell,c}[{k}] = {\bf h}_{\ell}^{\sf H}[{k}] {\bf f}_c +  \bar{z}_{\ell,c},
    \end{align}
    and ${\bf h}_{\ell}^{\sf H}[{k}] $ represents the ${k}$-th subchannel.
    This extension allows our technique to be applicable to wideband scenarios without significant changes.
      
    \subsection{Non-Stationary Channels}\label{Sec:Stationary}
    In the proposed beam alignment technique, we have assumed a stationary channel model along with the availability of sufficient beam training history for DNN training.
    This approach is effective for capturing all channel behaviors in stationary channels over time.
    However, we recognize that real-world scenarios often involve non-stationary channels, where channel conditions can change rapidly. In such cases, the DNN needs to adapt to the latest channel dynamics to maintain optimal performance. A practical approach is to implement online fine-tuning of the neural network using the most up-to-date results, ensuring that it continuously adjusts to the current channel conditions.
    Additionally, there are other options, such as reinforcement learning techniques like the multi-armed bandit (MAB) approach \cite{site_specific_MAB,BA_MAB_Lemma_derive}, to further enhance the DNN's output in dynamic channel environments. These online learning techniques can be integrated with the proposed technique to enhance its applicability in non-stationary channels and to improve its robustness and adaptability.
    
    \subsection{Other Sources of the Side Information}
    In the proposed beam alignment technique, we have considered only location information as side information when training the DNN to learn the beam prior probabilities. However, the applicability of the proposed technique is not limited to the availability of location information. Instead, it can be applied with a diverse range of side-information sources, including sub-6 GHz channel data \cite{sub6_1, sub6_2}, other relevant contextual information \cite{vision,Radar_beam}, and the incorporation of multi-modal data \cite{multi_modal_vision_position,multi_modal_magazine}. Even for these sources, the DNN can be effectively trained to establish a proper relationship between side-information and channel characteristics, enabling the proposed technique to be applied to various scenarios by utilizing relevant side-information effectively.

    \section{Simulation Results}\label{sec:Simulation}
    In this section, we evaluate the superiority of the proposed beam alignment technique over the existing techniques. In what follows, we first evaluate the performance under the assumption of perfect beam prior probability using a simplified channel model. We then conduct a performance evaluation using the realistic DeepMIMO channel dataset.
    
    
    \subsection{Evaluation with Perfect Beam Prior Probability}\label{sec:results_perfect}
    In this subsection, we evaluate the performance of the proposed technique under the assumption of perfect beam prior probability. 
    We simulate a dynamic channel environment where the BS, equipped with a 256-element ULA at half-wavelength spacing, conducts beam alignment.
    The scenario involves the BS located at the center of a cell, where the cell is divided into four distinct locations. Each location exhibits different beam prior probabilities influenced by obstructions and dynamic channel environments. As discussed in Sec.~\ref{sec:Analysis_beam_mis}, we consider perfect beam prior probabilities such that
    \begin{align}\label{eq:channel_assumption_sim}
        \mathbb{P}({\bf h}_{\ell} =  \alpha_{\ell} {\bf f}_c) =\hat{g}_{\ell,c},~\forall c\in\{1,\ldots,256\}. 
    \end{align}
    Specifically, we assume that the dynamic channels at each location are generated from four possible angles with the beam prior probabilities characterized in Table~\ref{table:beam_prior}.
    
    \begin{table}[t]
        \renewcommand{\arraystretch}{1.3}
        \caption{The beam prior probabilities associated with each location for simulations in Sec.~\ref{sec:results_perfect}.}\label{table:beam_prior}
        \small
        \centering
        \begin{tabular}{|c|c|c|c|c|c|} \hline
            \multicolumn{2}{|c|} {UE location $\ell$} & {1} & {2} & { 3} & {4} \\ \hline \hline
            \multicolumn{2}{|c|}{$\mathbb{P}({\bf h}_{\ell} = \alpha_{\ell} {\bf f}_1)$} & $0.7$ & $0.6$ & $0.5$ & $0.4$  \\ \hline 
            \multicolumn{2}{|c|}{$\mathbb{P}({\bf h}_{\ell} = \alpha_{\ell} {\bf f}_2)$} & $0.1$ & $0.2$ & $0.2$ & $0.3$  \\ \hline 
            \multicolumn{2}{|c|}{$\mathbb{P}({\bf h}_{\ell} = \alpha_{\ell} {\bf f}_3)$} & $0.1$ & $0.1$ & $0.2$ & $0.2$  \\ \hline 
            \multicolumn{2}{|c|}{$\mathbb{P}({\bf h}_{\ell} = \alpha_{\ell} {\bf f}_4)$} & $0.1$ & $0.1$ & $0.1$ & $0.1$  \\ \hline 
        \end{tabular}
    \end{table}

    We evaluate the beam misalignment probability with a fixed overhead of $256$ in each SNR range, where the SNR is denoted by ${\bf{h}_{\ell}^{\rm H}\bf{h}_{\ell}}/{\sigma^2}$.
    Our performance evaluation uses the Monte Carlo method, and each location is realized with equal probability by repeating the simulation $100,000$ times.
    For performance comparisons, we evaluate the following beam alignment techniques: (i) the exhaustive search, which searches over all the beam codewords with equal beam repetition as $r_{\ell, c} = \lfloor{R_{\rm sum}}/{N} \rfloor, \forall \ell,c$,
    and (ii) the top-$k$ search, in which the beam candidate set is fixed as $\mathcal{S}_{\ell}^{\star}(k)$, while the number of beam repetitions is fixed as $r_{\ell, c} = \lfloor{R_{\rm sum}}/{k} \rfloor, \forall \ell,c$. 
    
    \begin{table}[t]
        \renewcommand{\arraystretch}{1.4}
        \caption{The beam misalignment probability for various beam candidate sizes and SNR values at $\ell = 1$.}\label{table:Beam_mis_set_size}
        \setlength{\tabcolsep}{3pt}
        \footnotesize 
        \centering
        \begin{adjustbox}{max width=0.45\textwidth}
        {\begin{tabular}{|cc|cc|cc|cc|} \hline
            \multirow{1}{*}{} & \multicolumn{1}{|c|}{$S$} &\multicolumn{1}{|c|}{-18 dB}& \multicolumn{1}{|c|}{-16 dB}& \multicolumn{1}{|c|}{-14 dB} &\multicolumn{1}{|c|}{-12 dB} & \multicolumn{1}{|c|}{-10 dB} & \multicolumn{1}{|c|}{-8 dB} 
            \\ \hline \hline
            \multirow{4}{*}{Estimated}  & \multicolumn{1}{|c|}{1} & \multicolumn{1}{c|}{$\bf 0.3$} & \multicolumn{1}{c|}{$0.3$} & \multicolumn{1}{c|}{$0.3$} & \multicolumn{1}{c|}{$0.3$} & \multicolumn{1}{c|}{$0.3$} & \multicolumn{1}{c|}{$0.3$}  \\  \cline{2-8} 
            & \multicolumn{1}{|c|}{2} & \multicolumn{1}{c|}{$0.317$} & \multicolumn{1}{c|}{$\bf 0.272$} & \multicolumn{1}{c|}{$\bf 0.229$} & \multicolumn{1}{c|}{$0.207$} & \multicolumn{1}{c|}{$0.201$} & \multicolumn{1}{c|}{$0.200$} 
            \\  \cline{2-8} 
            & \multicolumn{1}{|c|}{3} & \multicolumn{1}{c|}{$0.444$} & \multicolumn{1}{c|}{$0.381$} & \multicolumn{1}{c|}{$0.263$} & \multicolumn{1}{c|}{$\bf 0.163$} & \multicolumn{1}{c|}{$0.113$} & \multicolumn{1}{c|}{$0.101$}
            \\ \cline{2-8} 
            & \multicolumn{1}{|c|}{4} & \multicolumn{1}{c|}{$0.608$} & \multicolumn{1}{c|}{$0.593$} & \multicolumn{1}{c|}{$ 0.410$} & \multicolumn{1}{c|}{$0.204$} & \multicolumn{1}{c|}{$\bf 0.064$} & \multicolumn{1}{c|}{$\bf 0.010$} 
            \\ \hline \hline
            \multirow{4}{*}{Simulated}  & \multicolumn{1}{|c|}{1} & \multicolumn{1}{c|}{$\bf 0.300$} & \multicolumn{1}{c|}{$0.300$} & \multicolumn{1}{c|}{$0.300$} & \multicolumn{1}{c|}{$0.300$} & \multicolumn{1}{c|}{$0.300$} & \multicolumn{1}{c|}{$0.300$}  \\  \cline{2-8} 
            & \multicolumn{1}{|c|}{2} & \multicolumn{1}{c|}{$0.317$} & \multicolumn{1}{c|}{$\bf 0.272$} & \multicolumn{1}{c|}{$\bf 0.229$} & \multicolumn{1}{c|}{$0.205$} & \multicolumn{1}{c|}{$0.199$} & \multicolumn{1}{c|}{$0.198$} 
            \\  \cline{2-8} 
            & \multicolumn{1}{|c|}{3} & \multicolumn{1}{c|}{$0.364$} & \multicolumn{1}{c|}{$0.317$} & \multicolumn{1}{c|}{$0.230$} & \multicolumn{1}{c|}{$0.153$} & \multicolumn{1}{c|}{$0.112$} & \multicolumn{1}{c|}{$0.101$}
            \\ \cline{2-8} 
            & \multicolumn{1}{|c|}{4} & \multicolumn{1}{c|}{$0.370$} & \multicolumn{1}{c|}{$0.360$} & \multicolumn{1}{c|}{$0.261$} & \multicolumn{1}{c|}{$\bf 0.145$} & \multicolumn{1}{c|}{$\bf 0.051$} & \multicolumn{1}{c|}{$\bf 0.008$}
            \\ \hline 
        \end{tabular}} 
        \end{adjustbox}
        \label{table:2}
    \end{table}

    In Table~\ref{table:Beam_mis_set_size}, we compare the estimated beam misalignment probabilities in \eqref{eq:upper_overall} with the simulated beam misalignment probabilities for various beam candidate sizes and SNR values at $\ell = 1$. 
    In Table~\ref{table:Beam_mis_set_size}, the minimum beam misalignment probability for each SNR is highlighted in bold. Despite some inherent error associated with the union bound, the results in Table \ref{table:Beam_mis_set_size} demonstrate that the estimated beam misalignment probabilities effectively guide the selection of an appropriate beam candidate set size.
    It is also shown that the estimated probability for a fixed SNR value has a single minimum point related to the beam candidate set size $S$. This observation justifies the stopping condition suggested in {\bf Algorithm 1}.
     
    
    \begin{figure}[t]
        \centering
        {\epsfig{file=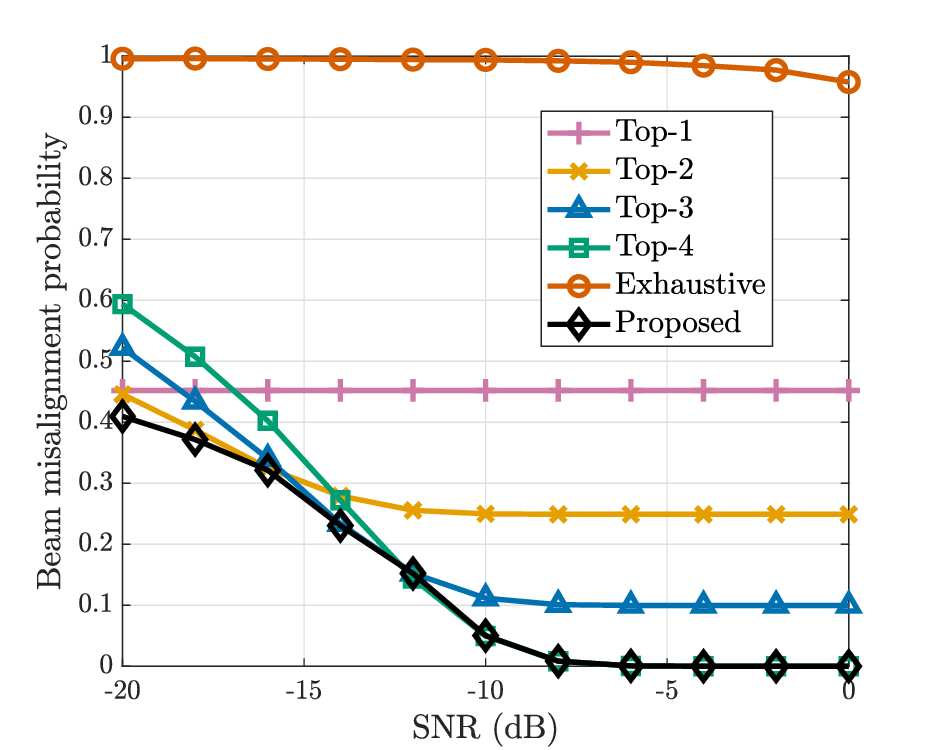,width=7.5cm}}
        \caption{Comparison of the beam misalignment probabilities of different beam alignment techniques for various SNR levels assuming perfect beam prior probability.} 
        \label{fig:Beam_mis_Synthetic}\vspace{-3mm}
    \end{figure}
    In Fig. \ref{fig:Beam_mis_Synthetic}, we compare the beam misalignment probabilities of different beam alignment techniques for various SNR levels. 
    Fig. \ref{fig:Beam_mis_Synthetic} shows that the proposed technique outperforms all existing techniques by adaptively optimizing both the beam candidate set ${\mathcal S}_{\ell}$ and the beam allocation vector ${\bf r}_{\ell}$ based on the SNR level. 
    In contrast, the top-$k$ search with a fixed value of $k$ performs well only in specific SNR regimes. 
    For instance, the top-$4$ search is inferior to the top-$2$ search in the low-SNR regime. This is because, in this regime, reducing beam determination errors becomes more crucial than increasing the likelihood of including the optimal beam in the candidate set.
    Conversely, the performance of the top-$2$ search is inferior to that of the top-$4$ search in the high-SNR regime and saturates to a non-zero value as the SNR increases. This is because, for $k=2$, the beam selection error remains constant, even though the beam determination error converges to zero as the SNR increases.
    
    \subsection{Evaluation with Realistic Channel Dataset}
    In this subsection, we evaluate the performance of the proposed technique using the DeepMIMO dataset generated by a realistic 3-D ray-tracing simulator \cite{deepmimo}.
    We specifically select the O$2$ dynamic (outdoor $2$) scenario at the $3.5$ GHz frequency band\footnote{Due to the lack of sufficiently dynamic and appropriate mmWave datasets, we adopted a dataset that operates at a carrier frequency of $3.5$ GHz for our experiments. However, considering that the channel's directionality is more dominant in mmWave environments, it is expected that the proposed technique will perform better in such scenarios.}. 
    This scenario considers a road with four car lanes (two in each direction) between the BS and UE, where the positions of the $50$ vehicles on the road change for each scene.
    This causes the wireless channel between the BS and UE to vary due to moving obstructions.
    The O$2$ dynamic scenario includes a total of $1,000$ dynamic scenes with $116,303$ users distributed across the grid. For simulation, we uniformly sample $100$ scenes of $961$ users from the first user grid (UG1) of this scenario. 
    Out of these $100$ scenes, we use the initial $80$ scenes for training the DNN and the remaining $20$ scenes for evaluation. 
    In our data generation setup, we consider BS with ULA configuration and an antenna spacing is set to half the wavelength. 
    During our simulation, all the location vectors are normalized and any nonzero channel vectors are removed. 
    The specific parameters for data generation are described in Table~\ref{table:deepmimo_parameter}.

    \begin{table}[t]
        \renewcommand{\arraystretch}{1.3}
        \caption{Key parameters and their values for the DeepMIMO dataset.}\label{table:deepmimo_parameter}
        \small
        \centering
        \begin{tabular}{|c|c|c|c|c|c|} \hline
            \multicolumn{2}{|c|}{\bf Parameter} & {\bf Value} \\ \hline \hline
            \multicolumn{2}{|c|}{Active BS} & $1$ \\ \hline
            \multicolumn{2}{|c|}{BS antenna} & $64 \times 1$ ULA \\ \hline
            \multicolumn{2}{|c|}{UE antenna} & Single\\ \hline
            \multicolumn{2}{|c|}{Antenna spacing} & $0.5$ \\ \hline
            \multicolumn{2}{|c|}{Bandwidth} & $500$ MHz \\ \hline
            \multicolumn{2}{|c|}{The number of OFDM subcarriers} & $512$ \\ \hline
            \multicolumn{2}{|c|}{OFDM sampling factor} & $1$ \\ \hline
            \multicolumn{2}{|c|}{OFDM limit} & $1$ \\ \hline
            \multicolumn{2}{|c|}{The number of channel paths} & $25$\\ \hline
        \end{tabular}
    \end{table}
    \begin{figure*}
        \centering 
        \begin{minipage}{2\columnwidth}
            \centering 
            {\setlength{\fboxrule}{0pt}
            \subfigure[$N=16$]
		{\epsfig{file=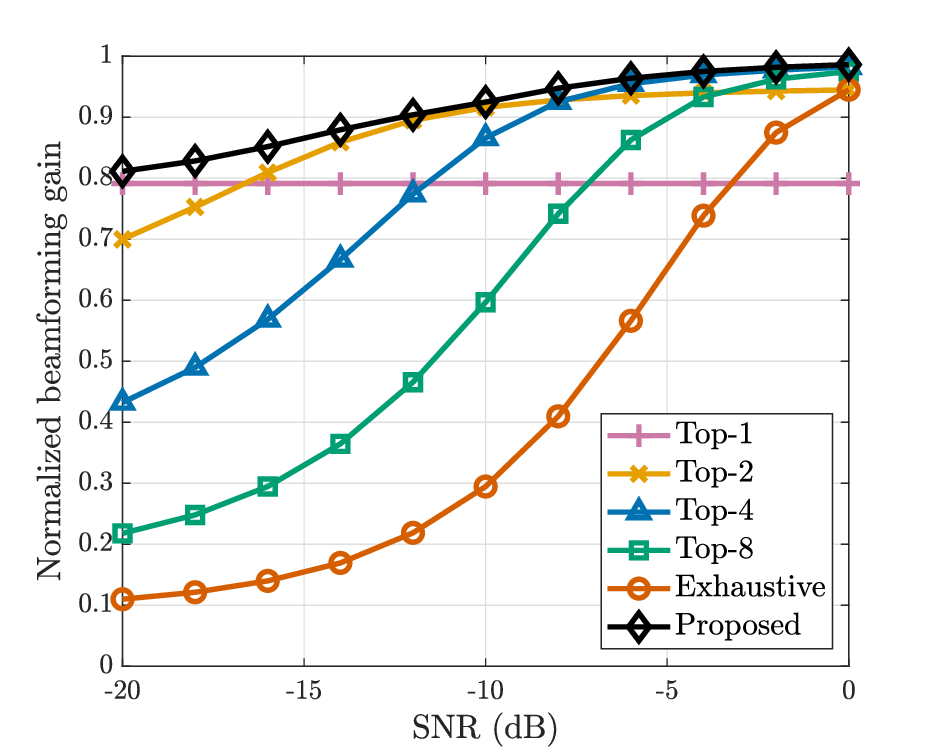, width=6.4cm}}\hspace*{-5.5mm}
            \subfigure[$N=64$]
		{\epsfig{file=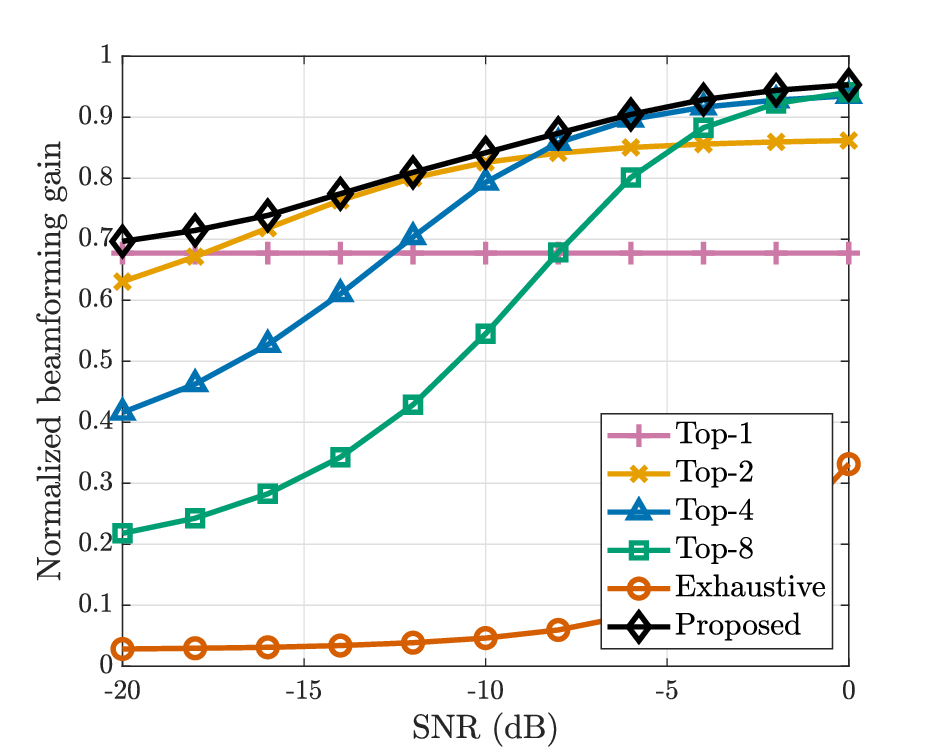, width=6.4cm}}\hspace*{-5.5mm}
            \subfigure[$N=256$]
            {\epsfig{file=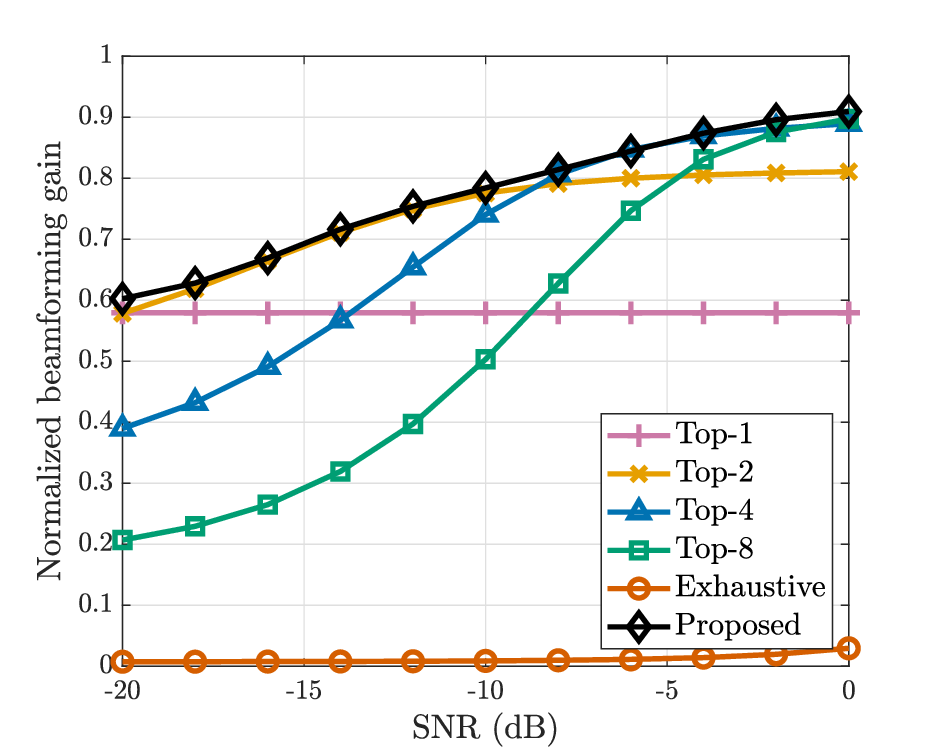, width=6.4cm}}\hspace*{-5.5mm}
            \caption{Comparison of the normalized beamforming gains of different beam alignment techniques for various SNR levels and antenna array sizes when $R_{\rm sum} = 256$.}
            \label{fig:diff_SNR}}
        \end{minipage}
    \end{figure*}
    
    To design the DNN for learning the beam prior probability vector, we adopt a fully-connected neural network that consists of $2$ input nodes, three hidden layers with $256$ nodes for each hidden layer, and an output layer with the same size as the number of antennas.
    The activation functions for the hidden layer and output layer are chosen as the ReLU and the softmax function, respectively. 
    We train the model using the ADAM optimizer with an initial learning rate of $0.01$ and set the number of epochs to $100$. Furthermore, because the output of the DNN is used to be probabilities, we apply a rough threshold of $1$ divided by the number of training scenes and set any output value lower than the threshold to $0$ during the operation process.
    Considering feedback issues in the proposed technique, we assume that the optimal beam repetition vector obtained from {\bf Algorithm~1} is compressed using linear regression as described in Sec. \ref{Sec:Feedback}. 
    As a performance metric, we consider the normalized beamforming gain defined as
    \begin{align}\label{eq:Beamforming_gain}
         G = \frac{|{\bf{h}}_{\ell}^{\rm H}{\bf{f}}_{\hat{c}_{\ell}^{\star}}|^2}{|{\bf{h}}_{\ell}^{\rm H}{\bf{f}}_{{c}_{\ell}^{\star}}|^2}.
    \end{align}
   


    In Fig.~\ref{fig:diff_SNR}, we compare the normalized beamforming gains of different beam alignment techniques for various SNR levels and antenna array sizes when $R_{\rm sum} = 256$.
    Fig.~\ref{fig:diff_SNR} shows that the beamforming gains of all techniques decrease as the number of antennas grows and the SNR decreases. Specifically, the exhaustive search experiences a significant performance drop, making it impractical for use in a large antenna array or in the low-SNR regime. Compared to the exhaustive search, both the proposed technique and the top-$k$ search show a performance gain by considering only the beam candidates with high beam prior probabilities based on the DNN. These results validate the efficacy of the beam prior information captured from the beam training history using the DNN. Although both the proposed technique and the top-$k$ search utilize the same beam prior information, the proposed technique exhibits consistently superior performance over the top-$k$ search regardless of the antenna array size and the SNR level.
    This implies that our optimization framework for the beam candidate set ${\mathcal S}_{\ell}$ and the beam allocation vector ${\bf r}_{\ell}$ is not only valid, but also crucial for maximizing the beamforming gain in a realistic scenario. 

    \begin{figure}[t]
        \centering
        {\epsfig{file=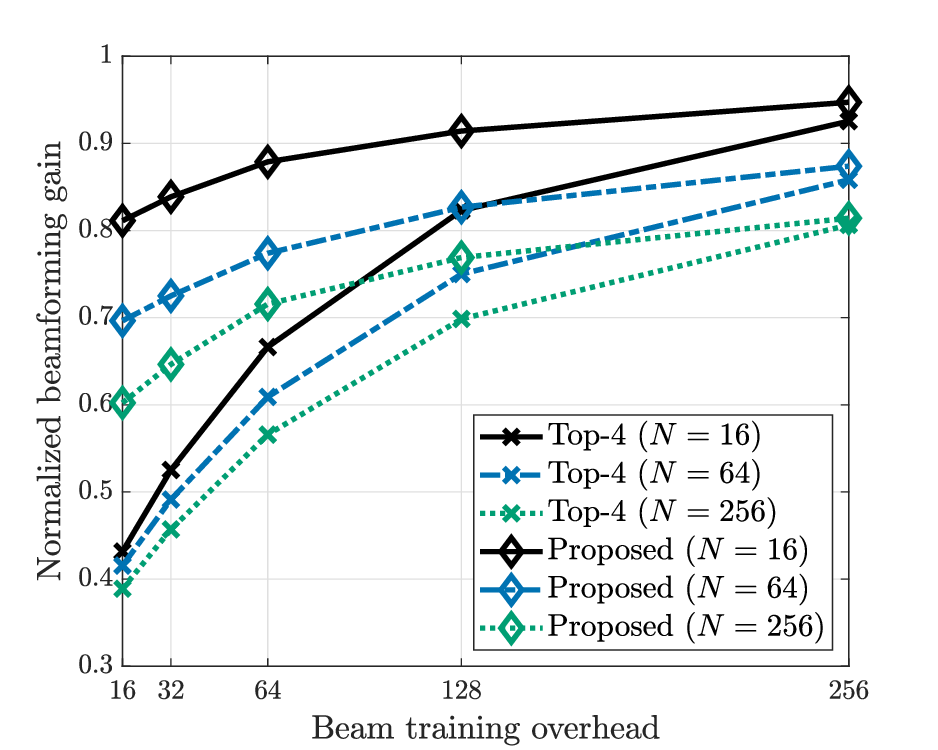,width=7.5cm}}
        \caption{Comparison of the normalized beamforming gains of different beam alignment techniques under different beam training overhead conditions at SNR of $-8$ dB.} 
        \label{fig:diff_R}\vspace{-3mm}
    \end{figure}

    In Fig.~\ref{fig:diff_R}, we compare the normalized beamforming gains of different beam alignment techniques under different beam training overhead conditions at SNR of $-8$ dB.
    Fig.~\ref{fig:diff_R} shows that the proposed technique exhibits robust performance even under conditions with a very limited number of beam training overheads. The observed trends under varying beam overheads are consistent with those across different antenna array sizes. In contrast, the top-$4$ search only performs well when sufficient overhead is provided, showing vulnerability under conditions with fewer beam training overheads. This is because the proposed technique adaptively decreases the beam candidate set size $S$ as the available beam training overhead decreases, indicating that our optimization framework successfully takes into account the effects of both SNR and the given beam training overhead. Furthermore, it is evident that performance compensation for low SNR is effectively attainable through beam training overhead, highlighting the adaptability and efficiency of the proposed technique in managing the trade-offs between SNR levels and beam training overhead.
    
    \begin{figure}[t]
        \centering
        {\epsfig{file=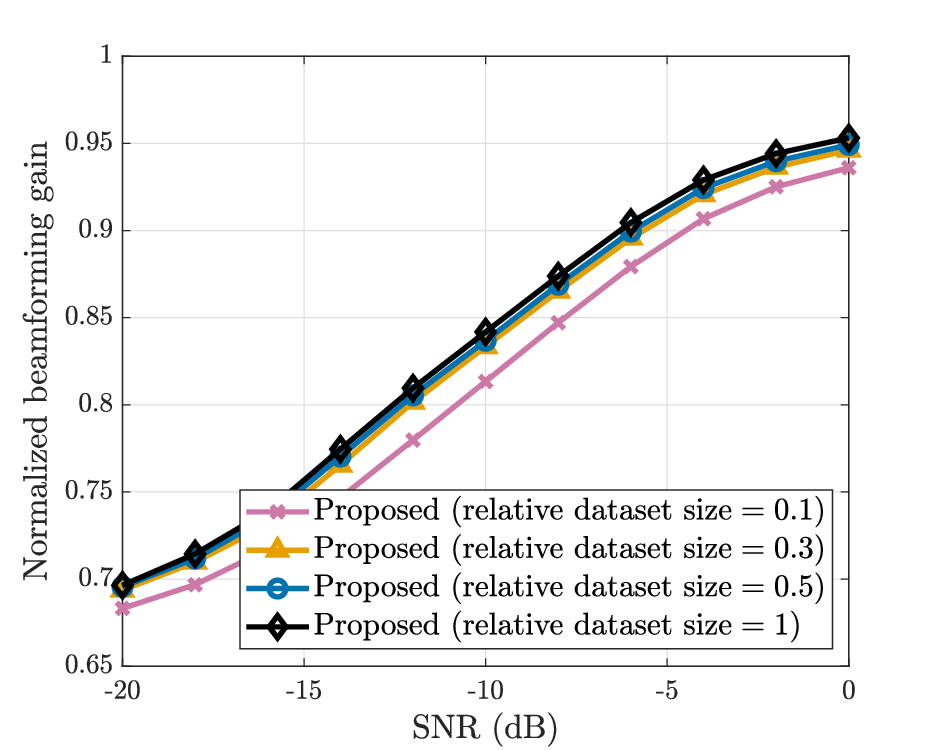,width=7.5cm}}
        \caption{The effect of the training dataset size on the normalized beamforming gain of the proposed beam alignment technique when $N=64$ and $R_{\rm sum} = 256$.} 
        \label{fig:diff_data}\vspace{-3mm}
    \end{figure}
    In Fig.~\ref{fig:diff_data}, we investigate the effect of the training dataset size on the normalized beamforming gain of the proposed beam alignment technique, specifically for configurations with $N=64$ and $R_{\rm sum}=256$. Fig.~\ref{fig:diff_data} shows that the performance of the proposed technique degrades as the size of the training dataset decreases, but the decrement is marginal. For instance, even with a 70\% reduction in the training dataset size, the performance degradation is approximately 1\% in terms of the normalized beamforming gain. This result demonstrates that the DNN adopted in the proposed technique can efficiently learn the relationship between location and optimal beam index, even with a relatively small training dataset obtained from beam training history.
    Although the overall performance degradation is marginal, a $10$-times reduction in the training dataset size results in a relatively large degradation. This trend is likely due to the challenges faced by the DNN when trained with limited data, particularly in comprehensively learning diverse beam histories for each location. In such cases, the use of reinforcement learning or online fine-tuning of the DNN would be crucial to improve the performance of the proposed technique, as discussed in Sec.~\ref{Sec:Stationary}.
    

    \begin{figure}[t]
        \centering
        {\epsfig{file=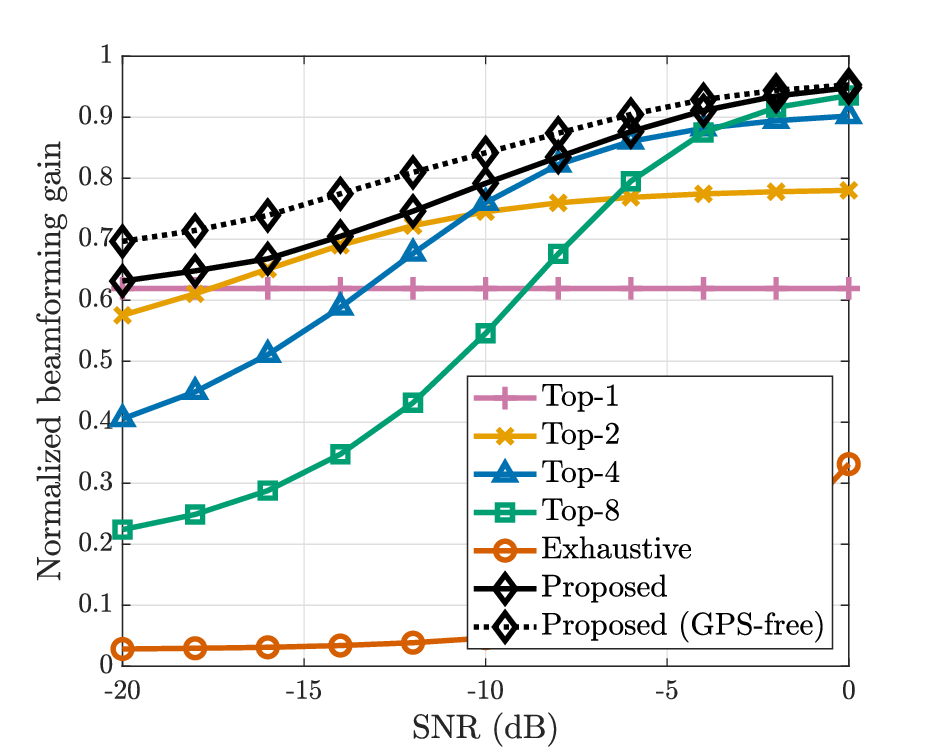,width=7.5cm}}
        \caption{The effect of the GPS error on the normalized beamforming gain of the proposed beam alignment technique when $N=64$ and $R_{\rm sum} = 256$.} 
        \label{fig:diff_SNR_GPS}\vspace{-3mm}
    \end{figure}
    In Fig.~\ref{fig:diff_SNR_GPS}, we investigate the effect of GPS error on the normalized beamforming gain of the proposed beam alignment technique for configurations with $N=64$ and $R_{\rm sum}=256$.
    Since the proposed technique leverages the location information of the UE, inaccuracies in the location coordinates, caused by outdated information from feedback delays and user mobility, might compromise the accuracy of the DNN output.
    To account for these inaccuracies, we follow the approach outlined in \cite{location_beam} using additive Gaussian noise with zero mean and a standard deviation of $2.04$ meters, which distorts the UE's Cartesian coordinates as $(\hat{x}_{\ell},\hat{y}_{\ell})=(\tilde{x}_{\ell}+n_{\mathrm x},\tilde{y}_{\ell}+n_{\mathrm y})$,
    where the noise components $n_{\mathrm x}$ and $n_{\mathrm y}$ follow $\mathcal{N}(0,2.04^2)$.
    Fig.~\ref{fig:diff_SNR_GPS} shows a general decline in performance for all the techniques that rely on beam prior probability, including the top-$k$ search and the proposed technique. 
    Although the exhaustive search does not suffer from the performance degradation due to the GPS error, it still shows the lowest beamforming gain compared to other techniques.
    The proposed technique not only consistently outperforms all other techniques, but also shows a larger performance gap with the top-$k$ search compared to scenarios without GPS errors in Fig.~\ref{fig:diff_SNR}. 
    It is also noticeable that the performance gap between the proposed technique with and without GPS errors decreases as the SNR increases. 
    These results indicate that our optimization framework remains effective, even in the presence of GPS inaccuracies. 


    \section{Conclusion}\label{sec:Conclusion}
    In this paper, we have proposed a novel beam alignment technique for low-SNR mmWave communications, which reduces the beam misalignment probability based on a beam repetition strategy. In particular, we have utilized the DNN to provide the information of beam prior probability at each location based on beam training history. Then, we have exploited the beam prior probability as channel priors to optimize both the selection of the beam candidates and the number of beam repetitions. We have demonstrated numerically that the proposed technique is an effective solution for enabling accurate beam alignment in dynamic low-SNR communication environments.

    An important direction for future research is to extend the proposed technique by utilizing various side-information sources along with a multi-modal approach.
    Another promising direction would involve applying online learning techniques, such as online fine-tuning or reinforcement learning. 

    \appendices
    \section{Proof of Lemma~\ref{lem:pair_wise_prob}}\label{apdx:pair_wise_prob}
    Given that $\bar{z}_{\ell,c} \sim \mathcal{CN}(0,\sigma^{2}/r_{\ell,c})$, the pair-wise miss-determination probability $\mathbb{P}\big( |\bar{z}_{\ell,j}|^2 > |{\alpha_{\ell}} + \bar{z}_{\ell,c}|^2\big)$ satisfies 
    \begin{align}
    \mathbb{P}\big( |\bar{z}_{\ell,j}|^2 > |{\alpha_{\ell}} + \bar{z}_{\ell,c}|^2\big) 
    =\mathbb{P}\Big(\frac{1}{r_{\ell,j}}\chi ^{2}_{2}(0) > \frac{1}{r_{\ell,c}} \chi ^{2}_{2}(\lambda_{\ell,c}) \Big),\nonumber
    \end{align}
    where $\chi ^{2}_{2}(\lambda_{\ell,c})$ represents a noncentral chi-squared random variable with 2 degrees of freedom and a noncentrality parameter $\lambda_{\ell,c} = 2 r_{\ell,c}{|\alpha_{\ell}|^2}/{\sigma^2}$.
    Recall that the probability density function of $\chi ^{2}_{2}(\lambda_{\ell,c})$ is given by
    \begin{align}\label{eq:apend_chi_pdf}
    f(x;\lambda_{\ell,c})=\frac {1}2e^{-\frac {1}2\left ({x+\lambda _{\ell,c}}\right)} \sum _{n=0}^\infty \frac {\lambda _{\ell,c}^{n}x^{n}}{4^{n}(n!)^{2}},\,x\ge 0, 
    \end{align}
    and $\chi ^{2}_{2}(0) $ follows an exponential distribution whose probability density function is given by
    \begin{align}\label{eq:apend_exp_pdf}
    g(y)= \frac {1}2 e^{-\frac {1}2y},\ y\ge 0. 
    \end{align}
    Since $\chi ^{2}_{2}(\lambda_{\ell,c})$ and $\chi ^{2}_{2}(0) $ are independent, the pair-wise miss-determination probability is rewritten as
    \begin{align}\label{eq:apend_ind}
    &\mathbb{P}\big( |\bar{z}_{\ell,j}|^2 > |{\alpha_{\ell}} + \bar{z}_{\ell,c}|^2\big) \nonumber \\
    &= \iint _{x\ge 0,y\ge 0,\frac{r_{\ell,j}}{r_{\ell,c}}x< y}f(x;\lambda_{\ell,c})g(y)dx dy \nonumber\\
    &=\int _{0}^\infty f(x;\lambda_{\ell,c})\int _{\frac{r_{\ell,j}}{r_{\ell,c}}x}^\infty g(y) dydx.
    \end{align}
    Similar to the derivation in \cite{BA_MAB_Lemma_derive}, applying \eqref{eq:apend_chi_pdf} and \eqref{eq:apend_exp_pdf} into \eqref{eq:apend_ind} yields
    \begin{align}
    &\mathbb{P}\big( |\bar{z}_{\ell,j}|^2 > |{\alpha_{\ell}} + \bar{z}_{\ell,c}|^2\big) \nonumber \\
    &=\int _{0}^\infty f(x;\lambda_{\ell,c})\int _{\frac{r_{\ell,j}}{r_{\ell,c}}x}^\infty \frac {1}2 e^{-\frac {1}2y} dydx\nonumber\\
    &=\int _{0}^\infty f(x;\lambda_{\ell,c}) e^{-\frac{r_{\ell,j}}{2r_{\ell,c}}x}dx \nonumber\\ 
    &=\frac {1}2e^{-\frac {1}2\lambda _{\ell,c}}\int _{0}^\infty e^{-\frac {1}2(1+\frac{r_{\ell,j}}{r_{\ell,c}})x}\sum _{n=0}^\infty \frac {\lambda _{\ell,c}^{n}x^{n}}{4^{n}(n!)^{2}}dx\nonumber\\
    &=\frac {1}2e^{-\frac {1}2\lambda _{\ell,c}}\sum _{n=0}^\infty \frac {\lambda _{\ell,c}^{n}}{4^{n}(n!)^{2}} \int _{0}^\infty e^{-\frac {1}2(1+\frac{r_{\ell,j}}{r_{\ell,c}})x}x^{n}dx\nonumber\\
    &=\frac {1}2e^{-\frac {1}2\lambda _{\ell,c}}\sum _{n=0}^\infty \frac {\lambda _{\ell,c}^{n}(\frac{2r_{\ell,c}}{r_{\ell,c}+r_{\ell,j}})^{n+1}}{4^{n}(n!)}\nonumber \\
    &=\frac{r_{\ell,c}}{r_{\ell,c}+r_{\ell,j}}e^{-\frac {1}2\lambda _{\ell,c}}e^{\frac{r_{\ell,c}}{2r_{\ell,c}+2r_{\ell,j}}\lambda_{\ell,c}}\nonumber\\
    &=\frac{r_{\ell,c}}{r_{\ell,c}+r_{\ell,j}}{\rm exp}\Big({-\frac{r_{\ell,c}r_{\ell,j}}{r_{\ell,c}+r_{\ell,j}}\rho_{\ell}}\Big),\nonumber
    \end{align}
    where $\rho_{\ell}=|\alpha_{\ell}|^2/\sigma^2$ is the received SNR at location $\ell$. This completes the proof.

    \section{Proof of theorem~\ref{thm:x_opt_S_2}}\label{apdx:x_opt_S_2} 
    
    The derivative of the beam miss-determination probability $\hat{p}_{{\rm miss-det},\ell}(x)$ with respect to $x$ is computed as
    \begin{align}\label{eq:p_miss_det_S_2_diff}
    &\! \! \frac{d \hat{p}_{{\rm miss-det},\ell}(x)}{d x} 
    \nonumber \\
    &=\big\{(\beta_{\ell}-k_{\ell}+1)x^2+(\beta_{\ell}-2)(k_{\ell}-1)x-(\beta_{\ell}+1)k_{\ell}+1\big\} \nonumber \\
    &\qquad \qquad \qquad \qquad \qquad \qquad \times  \frac{\hat{g}_{\ell,2}}{(x+1)^4}{\rm exp}{\left({-\frac{\beta_{\ell} x}{(x+1)^2}}\right)} \nonumber \\ 
    &={\big\{\underbrace{\beta_{\ell} (x+k_{\ell})(x-1)}_{\triangleq f(x)} - \underbrace{(k_{\ell}-1)(x+1)^2}_{\triangleq g(x)}\big\}} \nonumber \\
    &\qquad \qquad \qquad \qquad \qquad  \qquad \times \frac{\hat{g}_{\ell,2}}{(x+1)^4}{\rm exp}{\left({-\frac{\beta_{\ell} x}{(x+1)^2}}\right)} \nonumber \\ 
    &= h(x)\frac{\hat{g}_{\ell,2}}{(x+1)^4}{\rm exp}{\left({-\frac{\beta_{\ell} x}{(x+1)^2}}\right)},
    \end{align}
    where $h(x) \triangleq f(x)-g(x)$. 
    Since all the  terms in \eqref{eq:p_miss_det_S_2_diff}, except for $h(x)$, is positive for all $x$, we focus only on finding the point $x^\star > 0$ such that $h(x^\star) = f(x^\star)-g(x^\star)=0$. 
    One can easily see that $f(-k_{\ell})=f(1)=0$ and $g(-1)=0$, while $k_{\ell} > 1$ because $\hat{g}_{\ell,1} > \hat{g}_{\ell,2}$. 
    Consequently, over the interval $[-k_{\ell},1]$, $g(x)$ consistently exceeds $f(x)$, resulting in $h(x)$ taking on negative values. In other words, $\hat{p}_{{\rm miss-det},\ell} (x)$ decreases over the interval $(0,1)$.
    Further, for $x>1$, we have the following two cases: 
    \begin{enumerate}[(i)]
        \item {\bf Case 1:} If an intersection point exists between $f(x)$ and $g(x)$, and if $f(x)$ surpasses $g(x)$ from that point onwards, then the optimal minimum point $x^{\star}_{\ell}$ of $\hat{p}_{{\rm miss-det},\ell} (x)$ exists. 
        \item {\bf Case 2:} 
        If no such intersection point exists,
        then $f(x)$ is consistently less than or equal to $g(x)$, making $\hat{p}_{{\rm miss-det},\ell} (x)$ a decreasing function for $x > 0$, and the optimal $x^{\star}_{\ell}$ diverges to $\infty$.
    \end{enumerate} 
     It is worth noting that for $\beta_{\ell} \gg k_{\ell}-1$, the shape of $f(x)$ becomes sharper compared to $g(x)$, and the intersection point approaches $1$. This implies that as the SNR and beam training overhead increase, the solution converges towards $r_{\ell,1} \approx r_{\ell,2}$.
     It is also noticeable that as $k_{\ell}$ increases, the intersection point moves away from $1$ or ceases to exist. This implies that as the difference between the prior probabilities becomes larger, the optimal allocation tends to be $r_{\ell,1} \ll r_{\ell,2}$.

     In {\bf Case~1}, we have the following three scenarios:
      \begin{enumerate}[(i)]
        \item If $\beta_{\ell}>k_{\ell}-1$, then $h(x)$ forms a positive quadratic function with two distinct roots at $x>1$. The optimal minimum point can be calculated using the root-finding formula for $h(x)$:
        \begin{align}
        x^{\star}_{\ell} &= \frac{-a_{\ell,1}+\sqrt{a_{\ell,1}^2-4a_{\ell,2}a_{\ell,0}}}{2a_{\ell,2}}  \nonumber,
        \end{align}
        where $a_{\ell,2}=\beta_{\ell}-k_{\ell}+1$, $a_{\ell,1}=(\beta_{\ell}-2)(k_{\ell}-1)$, and $a_{\ell,0}=-(\beta_{\ell}+1)k_{\ell}+1$.
        \item If $\beta_{\ell}=k_{\ell}-1$ and $\beta_{\ell}>2$, then $h(x)$ forms a linear function with a positive slope, and the local minimum point can be calculated as $x^{\star}_{\ell} = -a_{\ell,0}/a_{\ell,1}$.
        \item If $2<\beta_{\ell}<k_{\ell}-1$, and $D=a_{\ell,1}^2-4a_{\ell,2}a_{\ell,0}>0$, then $h(x)$ forms a negative quadratic function with two distinct roots at $x>1$. The optimal minimum point can be calculated using the root-finding formula for $h(x)$:
        \begin{align}
        x^{\star}_{\ell} &= \frac{-a_{\ell,1}-\sqrt{a_{\ell,1}^2-4a_{\ell,2}a_{\ell,0}}}{2a_{\ell,2}}  \nonumber.
        \end{align}
    \end{enumerate}  
    In all other scenarios, which belong to {\bf Case 2}, the optimal $x^{\star}_{\ell}$ diverges to $\infty$. 
    This completes the proof.

    \bibliographystyle{IEEEtran}
    \bibliography{Reference}
    
\end{document}